\documentclass[
	aps,
	twocolumn,
	altaffilletter,
	nolongbibliography,
	numerical,
	flushbottom,
	secnumarabic,
	pra,
	superscriptaddress,
	floatfix,
	10pt
]{revtex4-2}

\usepackage{lipsum}
\usepackage{graphicx,subfigure}
\usepackage{braket}
\usepackage{amsmath, amssymb, amsthm}
\usepackage{newtxtext, newtxmath}
\usepackage{siunitx}
\usepackage{booktabs}

\usepackage[breaklinks, pdftex, hyperfootnotes=true, pdfpagelabels, bookmarks, pageanchor]{hyperref}
\hypersetup{%
	colorlinks=true, linktocpage=true, pdfstartpage=1, pdfstartview=FitH, pdfborder={0 0 0},%
	breaklinks=true, pdfpagemode=UseNone, pageanchor=true, pdfpagemode=UseOutlines,%
	plainpages=false, bookmarksnumbered, bookmarksopen=true, bookmarksopenlevel=1,%
	hypertexnames=true, pdfhighlight=/O,
	urlcolor=blue, linkcolor=blue, citecolor=blue, %pagecolor=RoyalBlue,%
}

\newcommand{\ie}    {i.\,e.}
\newcommand{\eg}    {e.\,g.}

\newcommand{\mingap}    {\Delta_\text{min}}

\DeclareMathOperator{\Tr}{Tr}
\DeclareMathOperator{\tts}{TTS}

\date{\today}
\begin{document}

\title{Standard quantum annealing outperforms adiabatic reverse annealing with decoherence}

\author{Gianluca Passarelli}
\email{gianluca.passarelli@spin.cnr.it}
\affiliation{CNR-SPIN, c/o Complesso di Monte S. Angelo, via Cinthia - 80126 - Napoli, Italy}

\author{Ka-Wa Yip}
\affiliation{Center for Quantum Information Science \& Technology, University of Southern California, Los Angeles, CA 90089, USA}
\affiliation{Department of Physics \& Astronomy, University of Southern California, Los Angeles, CA 90089, USA}

\author{Daniel A. Lidar}
\affiliation{Center for Quantum Information Science \& Technology, University of Southern California, Los Angeles, CA 90089, USA}
\affiliation{Departments of Electrical and Computer Engineering, Chemistry, and Physics, University of Southern California, Los Angeles, CA 90089, USA}

\author{Procolo Lucignano}
\affiliation{Dipartimento di Fisica “E. Pancini”, Università di Napoli Federico II}

\begin{abstract}

We study adiabatic reverse annealing (ARA) in an open system. In the closed system (unitary) setting, this annealing protocol allows avoidance of first-order quantum phase transitions of selected models, resulting in an exponential speedup compared with standard quantum annealing, provided that the initial state of the algorithm is close in Hamming distance to the target one. Here, we show that decoherence can significantly modify this conclusion: by resorting to the adiabatic master equation approach, we simulate the dynamics of the ferromagnetic $p$-spin model with $p=3$ under independent and collective dephasing. For both models of decoherence, we show that the performance of open system ARA is far less sensitive to the choice of the initial state than its unitary counterpart, and, most significantly, that open system ARA by and large loses its time to solution advantage compared to standard quantum annealing. These results suggest that 
%cast doubts concerning 
as a stand-alone strategy, ARA is unlikely to experimentally outperform standard ``forward'' quantum annealing, and that error mitigation strategies will likely be required in order to realize the benefits of ARA in 
%real-life applications.
realistic, noisy settings.

\end{abstract}

\maketitle

\section{Introduction}\label{sec:intro}

Quantum annealing (QA) was proposed more than two decades ago as a heuristic algorithm for finding ground states of Ising Hamiltonians $H_0 $~\cite{kadowaki:qa,Farhi:00} and since then applied to a variety of problems; see, for example, Refs.~\cite{Brooke1999,santoro-martonak,PhysRevE.70.057701,PhysRevE.71.066707,Matsuda:2009uq,Perdomo-Ortiz2012,Ramsey-expt,ronnow:speedup,Rieffel:2015aa,Azinovic:2016uq,Mott:2017aa,Li:comp-bio-2017,Mandra:2017ab,Jiang2018,Venturelli2019,Smelyanskiy:2018aa,Zlokapa:2019ab} and Refs.~\cite{Tanaka:book,albash:review-aqc,Hauke:2019aa} for recent reviews. In standard (forward) QA, a qubit system is initialized in the ground state of a transverse field driver Hamiltonian $V_\text{TF}$ (defined below) and evolved in time using the time-dependent Hamiltonian $H(t) = (1-s) V_\text{TF} + s H_0$, with $H_0$ a classical (Ising-type) Hamiltonian and the annealing schedule $ s = s(t) $ ramping from $s = 0$ to $s = 1$. In a closed system, provided $H=H(\theta)$ where $\theta=t/\tau$ and $\tau$ is the total evolution time, if $\tau$ is much longer than the adiabatic time scale $ \tau_{\text{ad}} =  |\bra{1}dH/d\theta\ket{0}|/\mingap^2 $, where $\mingap$ is the minimum of the gap $\Delta_{10}(s)$ between the instantaneous ground state $\ket{0(\theta)}$ and first excited state $\ket{1(\theta)}$, the system will remain in its instantaneous ground state at all times (for a more precise statement see, e.g., Refs.~\cite{Jansen:07,mozgunov2020quantum}). Hence, the final state will be the ground state of $H_0$, which may encode the solution to a classical optimization problem. In the open system setting of QA the evolution takes place in the presence of decoherence, and the associated timescales modify the statement of the adiabatic theorem~\cite{PhysRevA.71.012331,joye_general_2007,oreshkov_adiabatic_2010,Avron:2012mz}; in this case adiabatic evolution corresponds to remaining in the instantaneous steady state of the generator of the open system dynamics, and if the system Hamiltonian is swept at a rate that satisfies the adiabatic theorem for closed systems, the system will generally not end up in the ground state of $H_0$ at the end of the evolution~\cite{Venuti:2015kq}.

In recent years, many attempts have been made to improve upon the original QA algorithm~\cite{chancellor:reverse}. These include explicitly making use of diabatic transitions~\cite{Somma:2012kx,crosson2014different,brady_optimal_2021,venuti2021optimal,Crosson2021}, exploiting pauses to improve the success probability of the algorithm in the presence of decoherence~\cite{marshall,passarelli:pausing,PhysRevApplied.14.014100,albash2020comparing,izquierdo2020ferromagnetically}, diagonal catalysts~\cite{albash2021diagonal} and inhomogeneous driving to circumvent the problem of small spectral gaps around first-order quantum phase transitions (1QPTs)~\cite{nishimori:inhomogeneous-1,nishimori:inhomogeneous-2,PhysRevLett.123.120501,Adame:2020aa} (we do not discuss non-stoquastic approaches here; see, e.g., Refs.~\cite{Nishimori:2016aa,Albash:2019aa,Crosson2020designing}). Alternatively, in iterated reverse annealing (IRA)~\cite{yamashiro:ara} (originally called ``Sombrero adiabatic quantum computing''~\cite{perdomo:sombrero}), the system starts in a classical state at $s = 1$, then $s$ is decreased so as to increase the rate of quantum tunneling up to an inversion point $s_\text{inv}$, after which the annealing resumes as usual towards $s = 1$. In this variant of QA, quantum fluctuations are non-monotonic. With notable exceptions such as the adiabatic Grover algorithm~\cite{Roland:2002ul,RPL:10}, it is still an open problem whether standard forward annealing can provide a quantum advantage for optimization problems, and experiments with quantum annealing hardware have not been able to settle this beyond speedup results relative to particular algorithms such as classical simulated annealing~\cite{Albash:2017aa}. In contrast, weakly-decoherent IRA is presently believed to be a promising route towards quantum advantage~\cite{Crosson2021} and is actively being studied both theoretically and experimentally on D-Wave hardware~\cite{DWave-entanglement,Albash:2015pd,passarelli:reverse-ira,PhysRevApplied.15.014012,PhysRevA.104.012604,2020arXiv200708487K,Ikeda2019,Venturelli2019,bando2021breakdown}. IRA in the presence of decoherence was studied numerically in Ref.~\cite{passarelli:reverse-ira} for the $p$-spin model with $ p = 3 $ using the weak-coupling limit adiabatic master equation (AME)~\cite{zanardi:master-equations,albash:decoherence} unraveled using the time-dependent stochastic Monte Carlo wave function approach~\cite{yip:mcwf}. Ref.~\cite{passarelli:reverse-ira} showed that weak dephasing is the main driving mechanism for the enhancement of the ground state probability.
%%DL: this is too much information for the introduction:
%if the inversion point is chosen to be smaller than the location of the avoided crossing, $s_\text{min}$ [where $ \Delta(s_{\min}) = \Delta_{\min}$]. 
These findings have been confirmed experimentally in the $p=2$ case~\cite{bando2021breakdown}, and the role of spin bath polarization~\cite{lanting2020probing} beyond the weak-dephasing regime 
%in the description of open system effects 
in IRA was noted as well.

An alternative annealing protocol with non-monotonic quantum fluctuations is adiabatic reverse annealing (ARA), where instead of modifying the schedule $s = s(t)$, the system Hamiltonian is modified so as to enforce a classical initial condition via an additional term $H_\text{init}$~\cite{nishimori:reverse-pspin,yamashiro:ara} (we provide more details below). Systems that are subject to 1QPTs, such as the ferromagnetic $p$-spin model with $ p \ge 3 $, have been shown to benefit from the ARA protocol. In particular, Refs.~\cite{nishimori:reverse-pspin, yamashiro:ara} provide evidence that ARA allows avoidance of the exponentially closing gap associated with the 1QPT of this model, thus exponentially speeding up convergence to the ferromagnetic ground state relative to standard, forward QA. This result holds in the fully-coherent setting, when in addition the magnetization of the initial state is above a critical threshold~\cite{albash2021diagonal}. These promising theoretical closed system results motivate us to  undertake a critical examination of the effects of decoherence on the dynamics of ARA. Naturally, this aspect must be carefully addressed in order to understand the potential of ARA in realistic scenarios where decoherence is expected to be relevant.

To this end, here we study the effect of decoherence on ARA. We focus on the $p$-spin Hamiltonian with $p = 3$. This model is a tool that is commonly used to study the performance of quantum annealing~\cite{Jorg:2010qa} and we focus on it so as to extend the previous results of Refs.~\cite{nishimori:reverse-pspin, yamashiro:ara} from the closed to the open system setting. As shown in the next section, the permutational invariance of the $p$-spin model allows large instances of this model subject to collective dephasing to be simulated with a relatively mild computational effort. Independent dephasing is computationally more demanding and we aim to obtain some insights into this case by simulating smaller instances instead. We discuss these two models of dephasing below, for different choices of the model parameters, including the number of qubits $N$, transverse field strength, and initial magnetization.
 
%In a closed system, ARA is better than forward annealing in the $p$-spin model~\cite{yamashiro:ara}. In ~\cite{albash2021diagonal} they showed that it depends on other conditions, like the initial state. We show that in an open system these conditions can be relaxed, and there are certain degree of improvements due to open system decoherence. 
% It would also be good to consider examples where ARA is worse, and I know of one example (developed by Tameem Albash) where that is the case; this can be considered later.

As noted in Ref.~\cite{Crosson2021}, diabatic transitions to higher excited states may provide a shortcut towards the final target state. Therefore, we include in our studies annealing times $\tau$ shorter than the ones set by the adiabatic condition. We also explore cases with a small transverse field strength and a large Hamming distance between the initial state and the target state, which result in a very small and sharp gap and thus diabatic transitions to higher excited states. Therefore, we study the combined effect of decoherence and diabatic transitions.

The structure of this paper is as follows. In Sec.~\ref{sec:model}, we introduce the $p$-spin model and the ARA Hamiltonian. We additionally discuss the adiabatic master equation and the two dephasing models we consider. In Sec.~\ref{sec:spectrum}, we present the spectral properties of the $p$-spin Hamiltonian for several choices of the transverse field strength, initial magnetization, and number of qubits. This sets the stage for subsequent calculations. In Sec.~\ref{sec:dynamics}, we discuss the dynamical properties of this system during ARA for several choices of the Hamiltonian parameters and of the annealing time, in the presence of independent and collective dephasing. In Sec.~\ref{sec:tts}, we adopt the time to solution metric as a measure for the performance of ARA and compare unitary and weakly decohered ARA with standard QA. We present our conclusions in Sec.~\ref{sec:conclusions}.

\section{Model}\label{sec:model}

We focus on the ferromagnetic $p$-spin model. The number of spins (or qubits) is $N$. 

As opposed to standard QA, where the initial state is the ground state of the transverse field Hamiltonian (the state $\ket{+}^{\otimes N}$, where $\ket{+}=(\ket{0}+\ket{1})/\sqrt{2}$), in ARA the system is prepared in a classical configuration (a bitstring in the computational basis $\{\ket{0}\equiv \ket{\uparrow},\ket{1}\equiv \ket{\downarrow}\}$, the eigenbasis of the $\sigma^z$ Pauli operators with eigenvalues $\{+1,-1\}$, respectively). The Hamiltonian is purely longitudinal (i.e., the intensity of quantum fluctuations is zero) at $ t = 0 $ and at $ t = \tau $, where as above $ \tau $ is the total annealing time. The Hamiltonian depends on two time-dependent parameters, denoted $ s $ and $ \lambda $ in the following, and reads
\begin{equation}\label{eq:hamiltonian-ARA}
	H(s, \lambda) = s H_0 + (1 - s) (1 - \lambda) H_\text{init} + \Gamma (1 - s) \lambda V_\text{TF}.
\end{equation}
In this equation we set $s = \theta = t / \tau \in [0, 1]$, $\lambda \in [0, 1]$ satisfies $\lambda(s = 0) = 0$ and $\lambda(s = 1) = 1$, and the three operators $H_0$, $H_\text{init}$ and $V_\text{TF}$ are the target, initial, and transverse field Hamiltonians, respectively:
\begin{subequations}
	\begin{gather}
		H_0 = -J N\left(\frac{1}{N} \sum_{i = 1}^{N} \sigma^z_i \right)^p \qquad 2 \le p \le N,\label{eq:h0}\\
		H_\text{init} = -\sum_{i = 1}^N \epsilon_i \sigma^z_i,\label{eq:hinit}\\
		V_\text{TF} = -\sum_{i=1}^{N} \sigma^x_i.\label{eq:vtf}
	\end{gather}
\end{subequations}
In these equations, $ \sigma^{x,z}_i $ are Pauli operators acting on qubit $ i $. 
%The single-qubit computational basis is given by $ \sigma^z_i \ket{\sigma} = \sigma \ket{\sigma} $ with $ \sigma = \pm 1 $. 
%Sometimes in the following, for convenience, we will refer to the equivalent representation in terms of bitstrings, \ie, $\ket{0} \equiv \ket{\sigma = +1} \equiv \ket{\uparrow} $ and $\ket{1} \equiv \ket{\sigma = -1} \equiv \ket{\downarrow} $. 
The unit of energy is $ [J] = \SI{1}{\giga\hertz} $, thus the unit of time is $ [J]^{-1} = \SI{1}{\nano\second}$ (we set $ \hslash = 1 $ throughout). For every odd value of $p$, the target ground state of the $p$-spin model is the (non-degenerate) ferromagnetic state with all spins pointing up, \ie, $ \ket{\psi(s = 1)} = \ket{1, 1, \dots, 1} $. We choose $ p = 3 $ as it is the smallest value of $p$ for which the model undergoes a 1QPT in the thermodynamic limit when standard quantum annealing, corresponding to $\lambda = 1$, is considered~\cite{Jorg:2010qa}.

The Hamiltonian of Eq.~\eqref{eq:hinit} is diagonal in the computational basis and is used to set the initial state: $\ket{\psi(s = 0)} = \ket{\epsilon_1, \epsilon_2, \dots, \epsilon_N}$, where $\epsilon_i = \pm 1$ for all $i$.

The annealing path in ARA is specified by assigning the function $ \lambda = \lambda(s) $ in Eq.~\eqref{eq:hamiltonian-ARA}. We choose $\lambda(s) = s^q$ ($q > 0$) so that $s$ is the only free parameter and the evolution in the parameter space follows the path $(s, \lambda) \equiv (s, s^q)$ from $(0, 0)$ to $(1, 1)$. Additionally, we fix $q = 1$; standard QA is recovered by setting $q = 0$ (we recognize that $q = 1$ is a significant restriction; as shown in Ref.~\cite{nishimori:reverse-pspin,yamashiro:ara} this means that the 1QPT is avoided only for $c$ close to 1). As was proven in the thermodynamic limit and using the static approximation, ARA allows avoiding 1QPTs in the phase diagram of the $p$-spin model if the initial state is sufficiently close to the ferromagnetic ground state~\cite{yamashiro:ara}. The similarity between the initial and target states is expressed by the fraction of spin-up qubits $c = N_\uparrow / N$, which corresponds to a Hamming distance of $d_\text{H} = N (1 - c)$ and to an initial magnetization of $m_0 = \langle \sum_i \sigma_i^z \rangle_0 / N = 2c - 1$.

In a closed system, unitary setting, the analysis is simplified by noting that the Hamiltonian of Eq.~\eqref{eq:hamiltonian-ARA} is permutationally invariant. Hence, only the number of spin-up qubits ($ N_\uparrow $) and spin-down qubits ($ N_\downarrow $) is relevant but not their ordering. We can define total spin operators of the two subsystems, up and down, as in
\begin{equation}
	S_1^k = \frac{1}{2} \sum_{i=1}^{\lfloor N c\rfloor} \sigma_i^k, \quad S_2^k = \frac{1}{2} \sum_{i=\lfloor Nc\rfloor + 1}^{N} \sigma_i^k, \qquad k = x, y, z,
\end{equation}
where $ \lfloor x \rfloor $ is the largest integer smaller than $ x $. The Hamiltonian in Eq.~\eqref{eq:hamiltonian-ARA} commutes with the complete set of commuting operators $ \set{S_1^2, S_1^z, S_2^2, S_2^z} $, where $ S_j^2 = \vec{S_j} \cdot \vec{S_j} = S_{j,x}^2 + S_{j,y}^2 + S_{j,z}^2 $, with $ j = 1, 2 $. The three operators in Eq.~\eqref{eq:hamiltonian-ARA} are conveniently rewritten as
\begin{subequations}
	\begin{gather}
		H_0 = -N \left[\frac{2}{N}(S_1^z + S_2^z)\right]^p,\\
		H_\text{init} = -2 (S_1^z - S_2^z),\\
		V_\text{TF} = -2 (S_1^x + S_2^x).
	\end{gather}
\end{subequations}
The dynamics occur entirely in the tensor product of the two subspaces with maximum eigenvalues of $ S_1^2 $ and $ S_2^2 $, since these operators commute with the total Hamiltonian and the initial state lies in this subspace. Therefore, we can restrict numerical simulations to the subspace given by their tensor product, having dimension $ D = (\lfloor N c\rfloor + 1)(N - \lfloor Nc\rfloor + 1) $, quadratic rather than exponential in $N$. In the two subspaces, suitable bases are given by the simultaneous eigenstates of $ \set{S_j^2, S_j^z} $, denoted $ \ket{j; w} $ and such that $ S_j^z \ket{j; w} = (S_{j, \text{max}}^z - w) \ket{j; w} $, with $ S_{1, \text{max}}^z = \lfloor Nc\rfloor/2 $ and $ S_{2, \text{max}}^z = (N - \lfloor Nc\rfloor)/2 $. In this Dicke representation~\cite{Dicke:54}, the initial state is $ \ket{\psi(0)} = \ket{1; 0} \otimes \ket{2; N - \lfloor Nc\rfloor} $.

The interaction with the environment can change this picture. We employ the AME to describe the dynamics of the reduced density matrix $ \rho(s) $~\cite{zanardi:master-equations}, and unravel the master equation using the Monte Carlo wave function (MCWF) method~\cite{yip:mcwf}. We use the AME in the Lindblad form
\begin{equation}\label{eq:lindblad}
	\frac{1}{\tau} \partial_s \rho(s) = -i \left[ H(s, \lambda) + H_\text{LS}(s), \rho(s) \right] + \mathcal{D}\bigl[\rho(t)\bigr],
\end{equation}
where $H_\text{LS} $ is the Lamb shift and $ \mathcal{D} $ is the dissipator superoperator:
%\begin{gather}
\begin{align}
	H_\text{LS}(s) &= \sum_{\alpha, \beta} \sum_{a,b, a \ne b} S_{\alpha \beta}\bigl(\omega_{b a}(s)\bigr) L^\dagger_{\alpha a b}(s) L_{\beta a b}(s) \notag \\
	&\quad + \sum_{\alpha, \beta} \sum_{a,b} S_{\alpha \beta}(0) L^\dagger_{\alpha a a}(s) L_{\beta b b}(s)
\end{align}
\begin{align}
	\mathcal{D}\bigl[\rho(s)\bigr] &= \sum_{\alpha, \beta} \sum_{a, b, a \ne b} \gamma_{\alpha \beta}\bigl(\omega_{b a}(s)\bigr) \Bigl( L_{\beta a b}(s) \rho(s) L^\dagger_{\alpha a b}(s) \notag \\&\qquad-\frac{1}{2} \bigl\{ L^\dagger_{\alpha a b}(s) L_{\beta a b}(s), \rho(s) \bigr\} \Bigr) \notag \\
	&\quad \sum_{\alpha, \beta} \sum_{a, b} \gamma_{\alpha \beta}(0) \Bigl( L_{\beta a a}(s) \rho(s) L^\dagger_{\alpha b b}(s) \notag\\
	&\qquad-\frac{1}{2} \bigl\{ L^\dagger_{\alpha a a}(s) L_{\beta b b}(s), \rho(s) \bigr\} \Bigr).
\end{align}
%\end{gather}
%THERE IS SOMETHING TO FIX IN THIS EQ. DO WE REALLY NEED TO BE SO PEDANTIC AND SPLIT DIAGONAL AND NON DIAG TERMS?
In these equations, $ \omega_{b a}(s) = E_b(s) - E_a(s) $ are instantaneous Bohr frequencies [where $E_a(s)$ is the instantaneous eigenenergy of $H(s)$], $ L_{\alpha a b}(s) $ are Lindblad operators corresponding to $ \omega_{b a}(s)$, and $ \gamma_{\alpha \beta}(\omega)$ are relaxation rates [$ S_{\alpha \beta}(\omega) $ is their Hilbert transform]. Assuming the system-bath interaction Hamiltonian is given by 
\begin{equation}\label{eq:hamiltonian-sb}
	H_{SB} = g \sum_{\alpha} A_\alpha \otimes B_\alpha,
\end{equation}
where $ g $ is the coupling energy, $ A_\alpha $ are system operators and $ B_\alpha $ are bath operators, then the Lindblad operators are
\begin{equation}
L_{\alpha a b}(s) = \ket{E_a(s)} \braket{E_a(s) | A_\alpha | E_b(s)} \bra{E_b(s)} .
\end{equation}
We consider dephasing baths, and distinguish between two different kinds of dephasing~\cite{Duan:97,Zanardi:97c,Lidar:1998fk}:
\begin{enumerate}
	\item Collective dephasing: $ \alpha = 1 $ and there is only one coupling operator, \ie, $A = S^z_1 + S^z_2 =\sum_{i=1}^N \sigma^z_i$. All qubits are coupled to the same bath through the single operator $B$.
	\item Independent dephasing: $ A_\alpha $ are single-qubit operators $\sigma_i^z$, where $\alpha = i = 1,\dots , N$. Each qubit is coupled to its own independent bath.
\end{enumerate}
Case 2 breaks the permutation symmetry, while case 1 preserves it, hence allowing us to work in the subspace where the unitary dynamics occurs. 

In general, for long annealing times $ \tau $ the success probabilities given by simulation with the collective coupling assumption are higher than those given by the independent coupling.
In fact, the steady state solution $\rho(s = 1,\tau \rightarrow \infty)$ of the Lindblad master equation in the weak coupling limit is the Gibbs state set by the problem Hamiltonian:
\begin{equation}
	\rho(s = 1,\tau \rightarrow \infty) = \frac{e^{- \beta H_S(1)}}{Z} = \frac{e^{- \beta H_0}}{Z},
	\label{eq:rhoS0}
\end{equation}
where $Z = \Tr\left(\exp\left(- \beta H_0 \right)\right)$.

%For the collective dephasing model, the dynamics occurs entirely in the two subspaces with maximum eigenvalues of $S_1^2$ and $S^2_2$. The Lindbladian is defined according to the projection of $H(s)$ onto such  subspaces.

For the collective dephasing model, the steady state is given by
\begin{equation}
	\rho'(s = 1,\tau \rightarrow \infty) = \frac{e^{- \beta H_0'}}{Z'},
\end{equation}
where $Z' = \textrm{Tr}\left(\exp\left(- \beta H_0' \right)\right)$ and $H_0' = PH_0P$. Here $P = \sum_{\alpha} \ket{\psi_{\alpha}}\bra{\psi_{\alpha}}$ is the projection into the tensor product of the two subspaces with maximum eigenvalues of $S_1^2$ and $S^2_2$, and $\ket{\psi_{\alpha}} = \ket{1;w_1} \otimes \ket{2;w_2}$ is the basis that has the maximum eigenvalues of $S_1^2$ and $S_2^2$. The simple proof below shows that for long annealing times the success probability given by the collective dephasing model is always higher than the one given by the independent dephasing model.
\begin{proof}
	Since $H_0'$ is the projection of $H_0$ into the subspace with the largest values of $ S_1^2 $ and $ S_2^2 $, $Z' < Z$ ($Z$ is a sum over positive terms, while $Z'$ is a sum over a subset of these terms). The success (ground state) probability $\langle E_0 | \rho'(s = 1, \tau \rightarrow \infty) | E_0 \rangle = \frac{e^{- \beta E_0}}{Z'} > \frac{e^{- \beta E_0}}{Z} = \langle E_0 | \rho(s = 1, \tau \rightarrow \infty) | E_0 \rangle$, where $ E_0 $ is the ground state energy of both $H_0$ and $H'_0$. 
\end{proof}

In the following, we consider an Ohmic bath, where $ \gamma(\omega) $ (the indices $ \alpha $ and $ \beta $ can be dropped if all coupling constants are equal to each other) is given by
\begin{equation}\label{eq:ohmic}
	\gamma(\omega) = 2 \pi \eta \frac{\omega e^{-\lvert\omega\rvert/\omega_\text{c}}}{1 - e^{-\beta \omega}},
\end{equation}
where $ \eta $ is the dimensionless coupling strength, $ \omega_\text{c} $ is a high-frequency cutoff and $ \beta = 1/T $ is the inverse temperature (with $ k_\text{B} = 1 $). It satisfies the Kubo-Martin-Schwinger condition~\cite{kubo1957statistical, martin1959theory}. We fix $ T = \SI{12}{\milli\kelvin} = \SI{1.57}{\giga\hertz} $ and $ \omega_\text{c} = \SI{8\pi}{\giga\hertz} $. We use $ K = 5000 $ Monte Carlo trajectories in our MCWF simulations.

\section{Spectral properties of the ARA Hamiltonian}\label{sec:spectrum}

In this section we explore how the spectrum changes by tuning certain parameters of Eq.~\eqref{eq:hamiltonian-ARA}. This is important in the calculations of diabatic transition rates, excitation and relaxation rates, and thus the understanding of the open system behavior of ARA.

\begin{figure}[tb]
	\centering
	\includegraphics[width = 0.9\columnwidth]{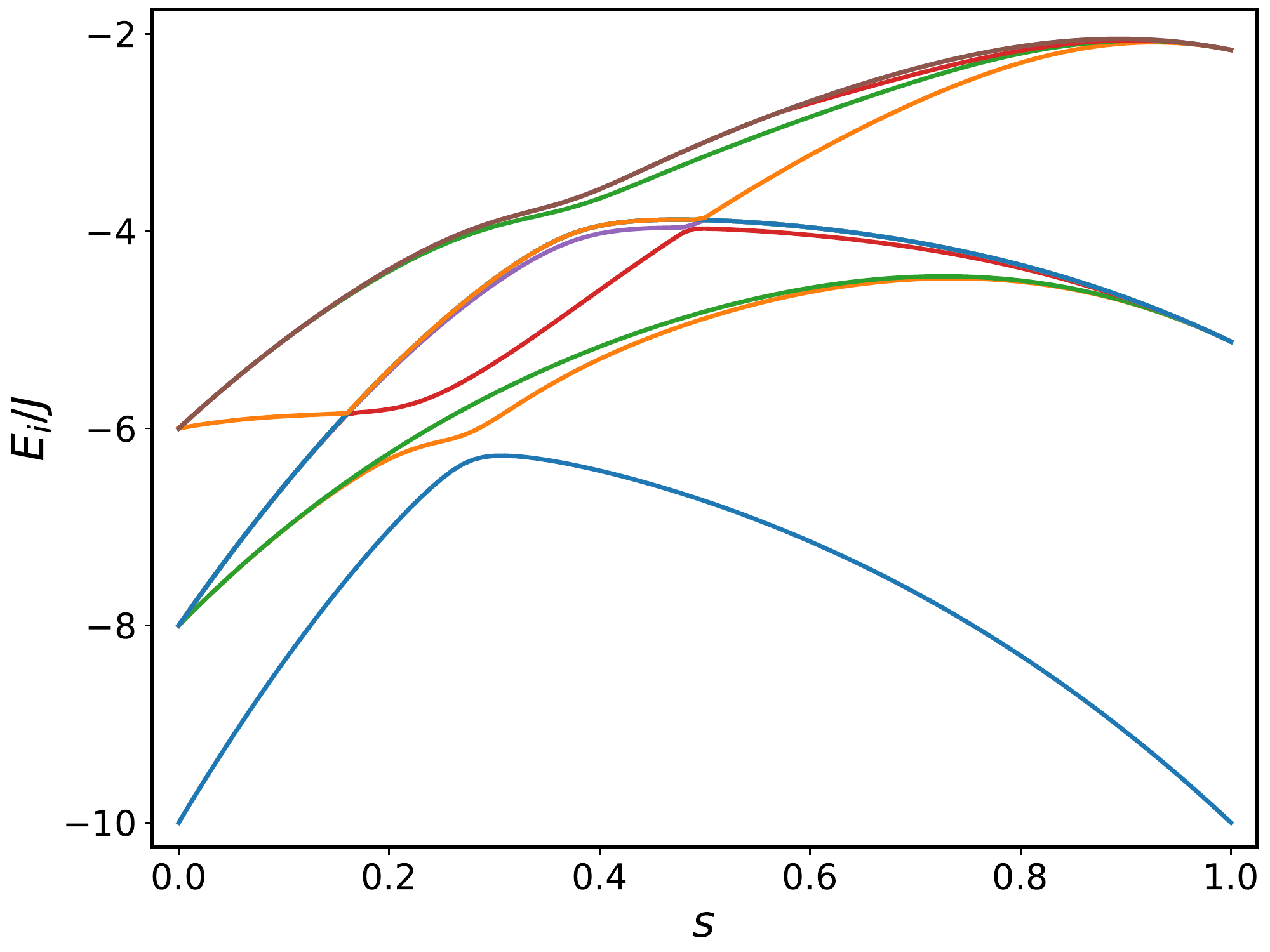}
	\caption{The $ 16 $ lowest-lying energies of an instance of the $p$-spin model with $ N = 10 $ qubits, with $H_{\text{init}} $ having $ c = 0.8 $.}
	\label{fig:ara-spectrum}
\end{figure}

Consider, for $N = 10$, the initial state with $N_\downarrow = 2$, so that $c = 0.8$. We plot the spectrum of the corresponding ARA Hamiltonian in Fig.~\ref{fig:ara-spectrum}.
The Bohr frequencues are $ \omega_{ij}(s) = E_{i}(s) - E_{j}(s)$. The corresponding minimum energy gaps are $\Delta_{ij} = \min_{s} \omega_{ij}(s)$. 

Similarly to standard QA, in ARA we are interested in the minimum energy gap $\Delta_{\min} = \Delta_{10}$ between the ground state and the first excited state:
The value of the gap is affected by the transverse field strength $ \Gamma $, the initial fraction $ c $ of spin-up qubits, and the number of qubits. In the following, we address all these dependencies for the parameters above.

\subsection{Dependence of the gap on the transverse field}

\begin{figure}[tb]
	\centering
	\includegraphics[width = 0.9\columnwidth]{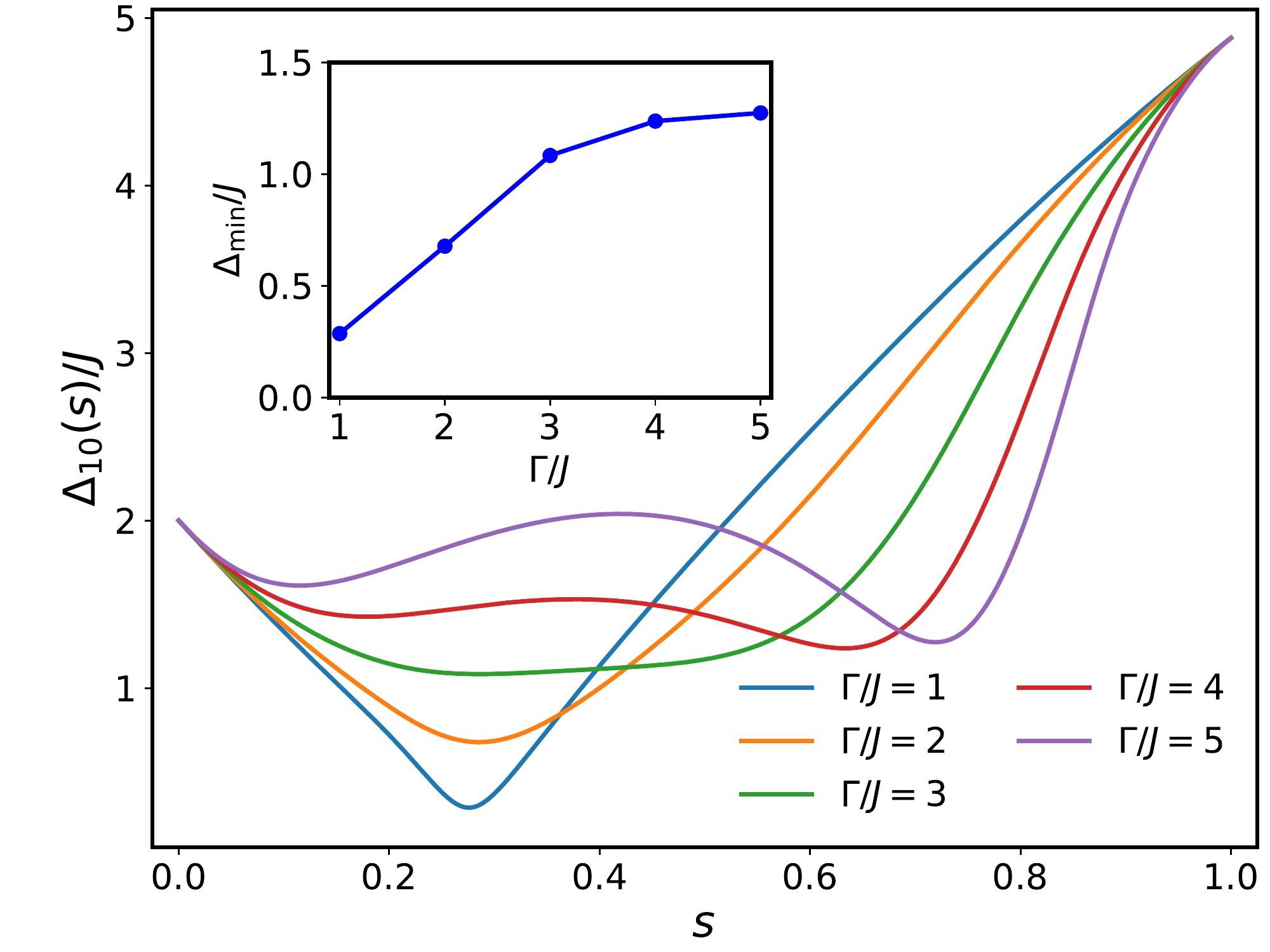}
	\caption{$\Delta(s)$ for $\Gamma/J \in \set{1, \cdots, 5}$ for $ N = 10 $ qubits and $ c = 0.8 $. In the inset, we show the minimum gap versus the strength of the transverse field.}
	\label{fig:nspin-10-mingap-gamma}
\end{figure}

%Let $ N = 10 $, the solution state of the $p$-spin problem is $\ket{1}^{\otimes 10}$. Consider the initial state with $ N_{\uparrow} = 8 $ ($ c = 0.8 $). We want to explore how the gap properties change with the transverse field $ \Gamma $ in Eq.~\eqref{eq:hamiltonian-ARA}.
We plot in Fig.~\ref{fig:nspin-10-mingap-gamma} the value of $\Delta_{10}(s)/J$, for $\Gamma/J \in \set{1, \cdots, 5}$. In the inset, we plot the dependence of $ \Delta_{\min}/J$ on $ \Gamma/J$. In general $\mingap$ decreases as $\Gamma$ decreases. The shape of the gap is sharper around the minimum for smaller values of $\Gamma$.

\subsection{Dependence of the gap on the fraction \texorpdfstring{$c$}{c}}

We wish to investigate how the gap properties change with different initial states, characterized by different values of $ c $. We focus here on $ N = 10 $ with $ \Gamma/J = 1 $, but the qualitative features obtained in this section are also found for other values of $ \Gamma $.

\begin{figure}[tb]
	\centering
	\includegraphics[width = 0.9\columnwidth]{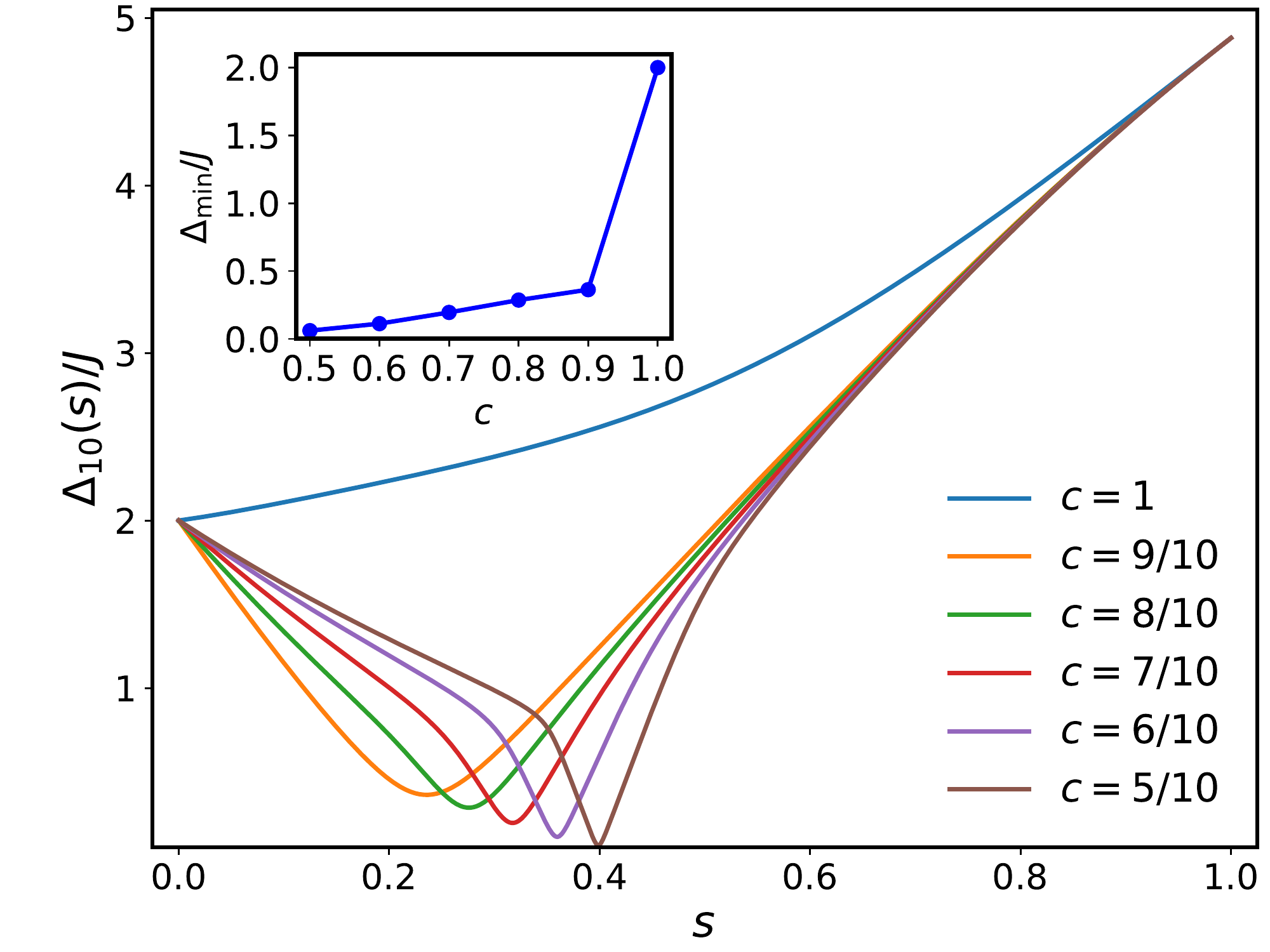}
	\caption{Instantaneous gap $ \Delta_{10}(s) $ for different values of the initial fraction $ c $ of up-aligned qubits, for $ N = 10 $ ($\Gamma/J = 1$).}
	\label{fig:nspin-10-mingap-c}
\end{figure}

We summarize our results in Fig.~\ref{fig:nspin-10-mingap-c}, from which we see that in general $\mingap$ decreases as the Hamming distance $d_\text{H} = N (1 - c)$ between the initial state and the target state increases.
%%DL: I don't think there is anything to be said about gap scaling here; scaling is as a function of N, and a fixed N result isn't informative about what happens in terms of the QPT either. See Fig.2 of Yamashiro et al. for the gap scaling with N.
%Quantitative differences in the scaling of the gap with $ c $ as a function of $ \Gamma $ (see the inset of Fig.~\ref{fig:nspin-10-mingap-c}) depend on whether or not the line $ \lambda(s) = s $ avoids the first-order quantum critical line in the phase diagram: if the 1QPT is avoided, as in this case, the scaling of the gap \DL{scaling with what?} is flatter, whereas in the presence of a 1QPT the decline is expected to be steeper. \DL{Where does the crossing happen for $N=10$? I find this last commentary unclear.}

\subsection{Scaling of the gap with the system size}

The dependence of the minimum gap on system size $N$ can be extrapolated to determine the adiabatic timescale for macroscopic systems. We study this dependence here for system sizes of $N\le 20$.

\begin{figure}[t]
	\centering
	\includegraphics[width = 0.9\linewidth]{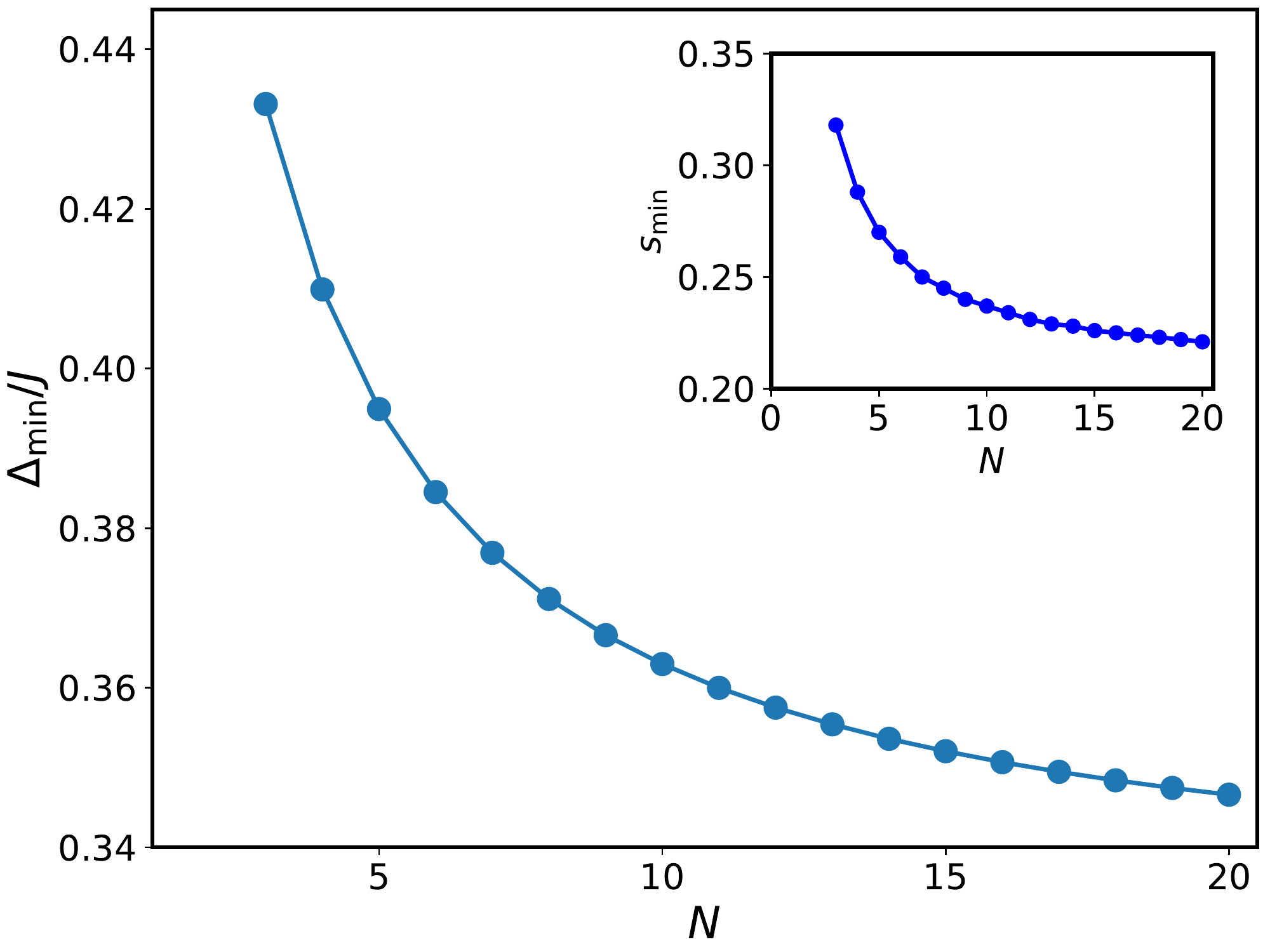}
	\caption{Minimum gap $\mingap$ for $ N \in \set{3, \cdots, 20}$. The initial state has $N_\downarrow = 1$. In the inset, we plot the value of $s_{\min}$ [Eq.~\eqref{eq:smin}] for each $N$.}
	\label{fig:mingap-vs-n}
\end{figure}

We first focus on the initial state with $ N_{\downarrow} = 1 $ (so that $ c = 1 - N_\downarrow / N $) and $ \Gamma/J = 1$, and plot $ \mingap $ for a range of $N$ values in Fig.~\ref{fig:mingap-vs-n}. In the inset, we also plot the annealing parameter value $ s_\text{min} $ where the minimum gap is found, \ie,
\begin{equation}
	s_{\min} = \arg\min_s \Delta_{10}(s).
	\label{eq:smin}
\end{equation}
This is important for the calculation of relaxation rates for ARA.

\begin{figure}[tb]
	\centering
	\includegraphics[width = 0.9\columnwidth]{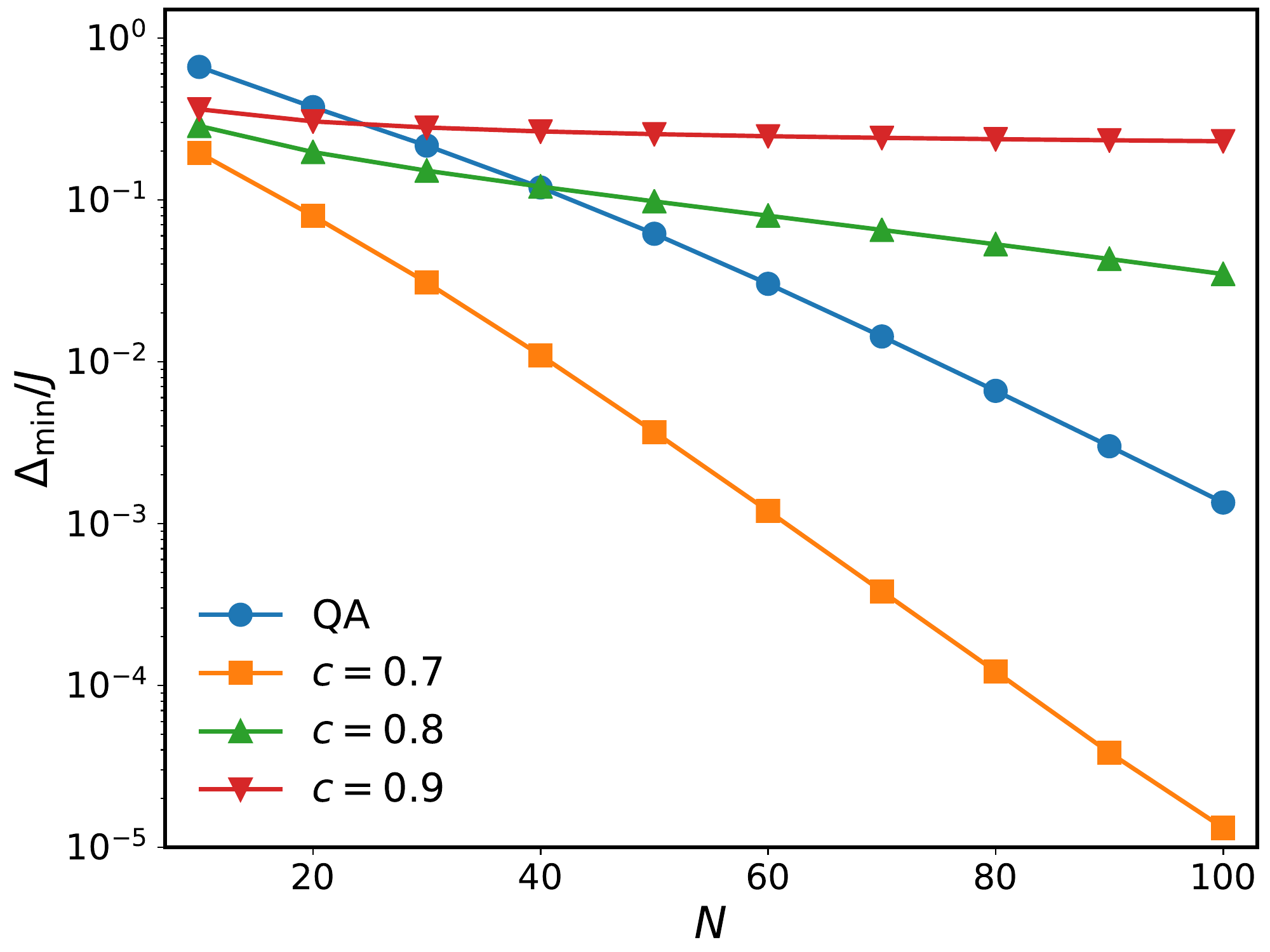}
	\caption{Minimum gap $ \mingap $ as a function of the system size $N$ for different values of $c$ ($\Gamma/J = 1 $). QA denotes standard forward annealing.}
	\label{fig:minimal-gap}
\end{figure}

Alternatively, one can fix the initial fraction $ c $ and study the scaling of $ \mingap $ as a function of $N$. To this end, it is more convenient to work in the symmetric sectors so as to study large systems and infer the behavior of the gap in the thermodynamic limit. In Fig.~\ref{fig:minimal-gap}, we report the scaling of the minimum gap as a function of $N$ for $ \Gamma / J = 1 $. In standard quantum annealing, the scaling of $ \mingap $ as a function of  system size is exponential for $p \ge 2$. As is clear from the figure, the scaling is exponential also for ARA when $ c $ is below a certain threshold, \eg, $ c < 0.9 $. In contrast, for $c \ge 0.9$  the gap is nearly constant over the range of system sizes we have considered (we expect it to  decrease as an inverse polynomial since the system traverses a second order QPT) as a function of system size. These results are in agreement with Ref.~\cite{yamashiro:ara}.

\section{Open system dynamics}\label{sec:dynamics}

In this section we study the open system, dephasing dynamics of the $p$-spin model with $p = 3$. We  discuss independent and collective dephasing and compare ARA with standard quantum annealing. 
More specifically, we explore four general classes of problems:
\begin{enumerate}
	\item Adiabatic reverse annealing in an open system 
($\mathsf{ARA_\text{Open}}$);
	\item Adiabatic reverse annealing in a closed system ($\mathsf{ARA_\text{Closed}}$);
	\item Standard quantum annealing in an open system ($\mathsf{QA_\text{Open}}$);
	\item Standard quantum annealing in a closed system ($\mathsf{QA_\text{Closed}}$).
\end{enumerate}
The comparison between $\mathsf{ARA_\text{Closed}}$ and $\mathsf{QA_\text{Closed}}$ was made in Ref.~\cite{yamashiro:ara} and it was shown that $\mathsf{ARA_\text{Closed}}$ can outperform $\mathsf{QA_\text{Closed}}$. Earlier studies (\eg, Refs.~\cite{TAQC,arceci:owp,theis2018gapindependent,passarelli:reverse-ira,passarelli:pspin}) already showed that decoherence can improve quantum annealing. Our goal here is to investigate whether and under which conditions such an improvement may be expected for $\mathsf{ARA_\text{Open}}$ \textit{vs} $\mathsf{ARA_\text{Closed}}$. More specifically, we wish to determine whether $\mathsf{ARA_\text{Open}}$ has any computational advantage over $\mathsf{QA_\text{Open}}$, which would support the use of the former protocol in a realistic experimental setting. We quantify  advantage/enhancement in terms of two success metrics: the success probability $ p_\text{g} $ and the time to solution (TTS) (for alternative metrics see, \eg, Ref.~\cite{Vinci:2016tg}).

The success probability $ p_\text{g} $ is the probability of the final state being the target solution state, \ie, in our case:
\begin{equation}
	p_\text{g}(\tau) = \bra{1 \cdots 1}\rho(\tau)\ket{1 \cdots 1}.
\end{equation}

\begin{figure*}[t]
	\centering
	\includegraphics[width = 0.9\textwidth] {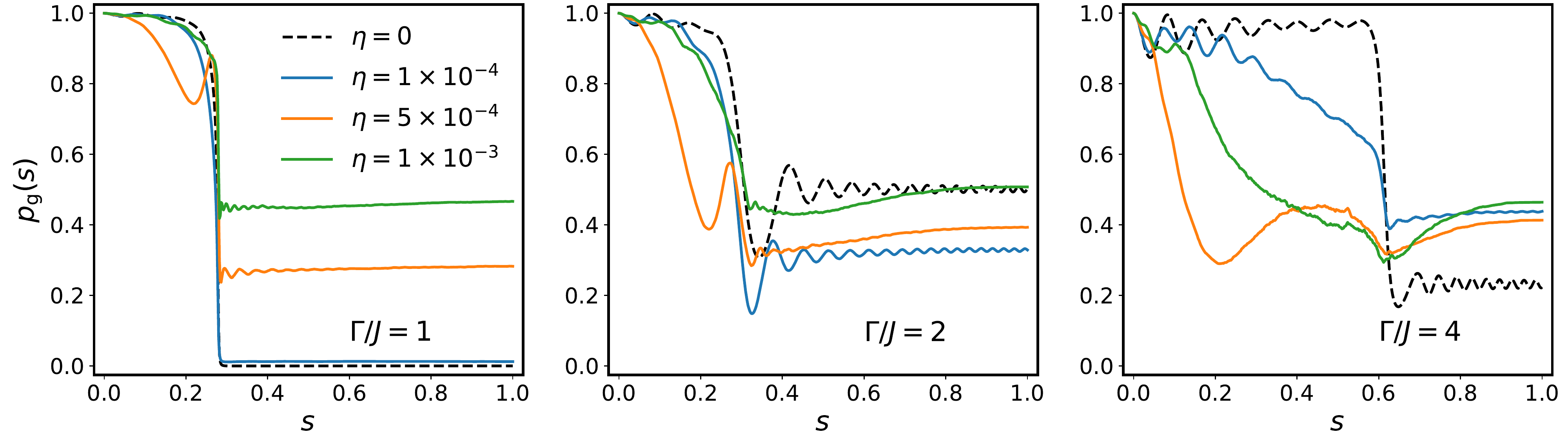}
	\caption{Ground state probability for collective dephasing as a function of $s$, for $\Gamma/J = \text{\numlist{1;2;4}}$ and several values of the coupling strength ($\eta = \text{\numlist{1e-4;5e-4;1e-3}}$). Other parameters are: $N = 50 $, $ c = 0.8 $, $ J\tau = 40$.}
	\label{fig:pgs-collective}
\end{figure*}

The time to solution (TTS) is defined as
\begin{equation}\label{eq:tts}
	\tts(\tau, p_\text{d}) = \tau \frac{\log(1 - p_\text{d})}{\log p_\text{e}(\tau)},
\end{equation}
where $ p_\text{e}(\tau) = 1 - p_\text{g}(\tau) $ is the error probability (\ie, the probability of ending up in an excited state at $ s = 1 $) and $ p_\text{d} $ is a threshold probability. The TTS represents the effective time it takes to solve the given problem at least once with a probability greater than $ p_\text{d} $ using multiple runs of duration $ \tau $~\cite{q108}. We fix $ p_\text{d} = 0.99 $. 

For closed systems, the TTS has different behaviors depending on the value of the annealing time $\tau$. Namely, in the quench limit, \ie, when $\tau $ is so short that the final ground state probability is effectively zero, the TTS is expected to diverge. If $ \tau $ is longer but not yet adiabatic, the error probability is described by the Landau-Zener formula $ p_\text{e}(\tau) = \exp(-\pi \tau / 4 \tau_\text{ad})$.
%%DL: This definition is dimensionless. The definition with units of time is given in the introduction.
%where $ \tau_\text{ad} = \max_t \lvert{\dot{H}(t)}_{10}\rvert/\Delta_{10}^2(t) $.
In this regime, the TTS has a plateau $ \tts^* = -4 \tau_\text{ad} \log(1-p_\text{d})/\pi $, which provides a measure of $\tau_\text{ad} $. Finally, in the adiabatic regime, the final time error is typically a power law of the annealing time, $ p_\text{e} \sim (\tau/\tau_\text{ad})^{-2} $, thus the time to solution scales as $ \tts \sim \tau / \log\tau $. We proceed to show that the presence of dephasing can affect these behaviors.

\subsection{Collective dephasing}

We study the dynamics of a system of $ N = 50 $ qubits and fix $ J\tau = 40 $. We also fix $ c = 0.8 $ so that $ N_\uparrow = 40 $. We vary the strength of the transverse field ($\Gamma/J \in \{ 1,2,4\}$) and the coupling to the environment ($\eta \in \{10^{-4},5\times 10^{-4},10^{-3}\}$).
The instantaneous ground state probability, $ p_g(s) = \bra{E_0(s)} \rho(s) \ket{E_0(s)} $, is plotted in Fig.~\ref{fig:pgs-collective}, in which we also report the closed system result ($\eta = 0$, or $\mathsf{ARA_\text{Closed}}$) for comparison.

The most significant aspect seen in Fig.~\ref{fig:pgs-collective} is that the case with the strongest system-bath coupling ($\eta = 10^{-3}$) results in the highest success probability at $s=1$. The unitary dynamics $\eta=0$ is quite sensitive to the value of $\Gamma/J$ and in two cases ($\Gamma/J = 1,4$) the open system dynamics results in a higher final success probability than the closed system dynamics.

In more detail, for $ \Gamma/J = 1 $, we see that the environment favors the ferromagnetic alignment and the ground state probability at $ s = 1 $ increases with respect to $\mathsf{ARA_\text{Closed}}$, in agreement with known findings concerning decoherence-assisted quantum annealing~\cite{passarelli:pspin,passarelli:pausing,passarelli:reverse-ira}.

For $ \Gamma/J = 2 $, on the other hand, dephasing reduces the success probability for the two smaller values of $\eta$ compared to the unitary case; this is due to the fact that $ p_\text{g} $ is already large for $\mathsf{ARA_\text{Closed}}$.
%, therefore excitations from the ground state are more likely than relaxations to the ground state. Overall, the curves are very similar to each other and the impact of decoherence appears to be marginal. Of course, this may change for longer annealing times, which will be discussed in Sec.~\ref{sec:tts}. 

For $\Gamma/J = 4$, we observe that the success probability is again increased compared to $\mathsf{ARA_\text{Closed}}$.
%, as opposed to the case of $ \Gamma/J = 2 $. 
Incidentally, this is the only case where the dependency of $p_\text{g}$ as a function of $ \eta \ne 0 $ is non-monotonic.

The behavior of the closed system curves is easily understood in terms of the instantaneous gap. Among the three cases, $ \Gamma/J = 2 $ has the largest minimum gap and thus the largest $p_\text{g}$. The coherent oscillations for $ \Gamma/J = 4 $ around $ s = 0 $ is explained by noting that the instantaneous gap is slowly varying in this region, and the frequency of oscillations ($ \num{\approx 1.9}J $) is close to that of Rabi oscillations between the two assumed constant lowest-lying energy states (where $ E_1 - E_0 \approx 1.5 J $).

At $t = 0$, the gap between the ground and first excited state is $\Delta_{10}(t = 0) = 2J$ and is independent of $\Gamma$. We note that the temperature in open system simulations ($T = \SI{12}{\milli\kelvin} = 1.57 J $) is comparable with the gap at $ t = 0 $, thus we expect thermal processes to be relevant already at the beginning of the dynamics. As shown in Figs.~\ref{fig:nspin-10-mingap-gamma} and~\ref{fig:nspin-10-mingap-c}, the instantaneous gap decreases as the annealing fraction $ s $ increases (with possible nonmonotonic behaviors for large values of $ \Gamma $), up to the minimum gap $\mingap$. The thermal energy scale remains comparable with the instantaneous gap up to a certain annealing fraction $s^*$, which depends on $\Gamma$. Then, the gap increases towards the value $ \Delta_p = J N (1-(1-2/N)^p) $ at $s = 1$, i.\,e., the gap of the $p$-spin model. For large system sizes, $\Delta_p \to 2 J p$. For $ p = 3 $, the final gap for finite-size systems is smaller than $\Delta_3 = 6 J$ and the thermal energy scale is comparable with the gap for the entire evolution. Thus, thermal effects are non-negligible up to $ s = 1 $. 

%Concerning the open system dynamics, we note that the temperature energy scale of $ T = \SI{12}{\milli\kelvin} = \SI{1.57}{\giga\hertz} $ is comparable with the gap already at $ s = 0 $ and up to a certain $ s = s^* $ after the minimum gap. For $ \Gamma/J = 1 $ we find $ s^* = 0.45 $, while for $ \Gamma/J = \text{\numlist{2;4}} $ we find $ s^* = 0.65 $. This is why, towards the end of the dynamics, the environment no longer significantly affects the ground state probability. \DL{This paragraph isn't quite clear. Numbers are needed to compare temperature to the gap sizes (what are the latter?) and the final sentence is a bit mysterious.} 

\subsection{Independent dephasing}

We now turn our attention to the independent dephasing case. As mentioned above, this case breaks the spin symmetry, which limits our simulations to few qubit systems. We set the coupling to {$ \eta = \num{1e-4} $} and the cutoff frequency to  {$ \omega_\text{c} = 8\pi\, \si{\giga\hertz} $}.

\subsubsection{$ N = 8 $, $ \Gamma/J = 1 $, $ J\tau = 250 $} 
\begin{figure}[t]
	\centering
	\includegraphics[width = 0.9\columnwidth]{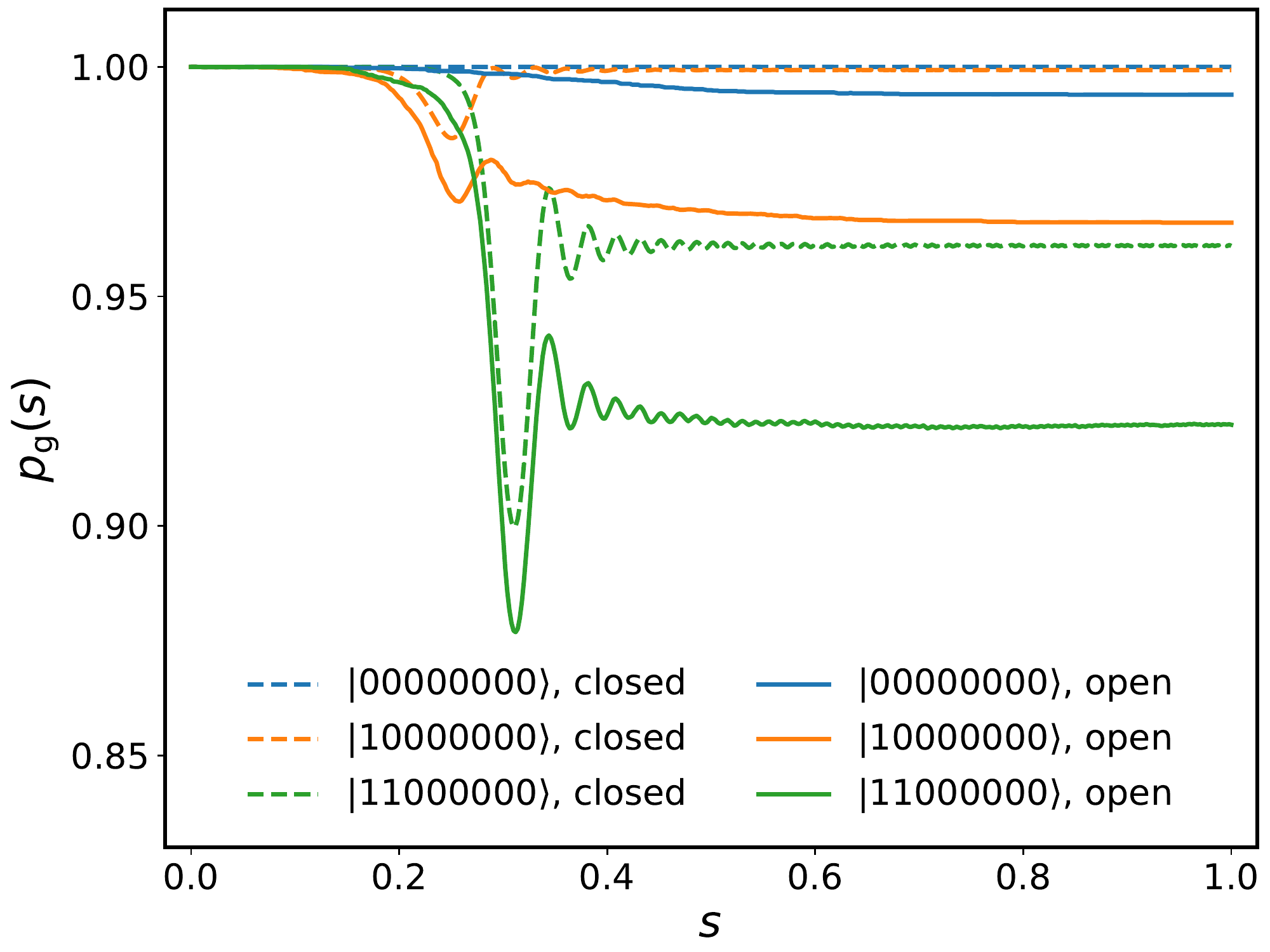}
	\caption{Open system and closed system ARA simulation results. Parameters are $ N = 8 $,  $ J\tau = 250 $, $ \Gamma/J = 1 $.}
	\label{fig:8ame_727}
\end{figure}

The results for closed and open system ARA, starting from different initial states,  are plotted in Fig.~\ref{fig:8ame_727}. 
%The solid lines are open system simulation results coming from the individual dephasing model while the dashed lines are results of closed system simulations.

If the initial state is already the target ferromagnetic state ($ c = 1 $), then we see that the ground state probability is close to one for the entire dynamics both in the unitary and the open system case. The spectral gap $ \Delta_{10}(s) $ is always larger than the temperature energy scale, thus thermal excitations are unlikely. Moreover, the gap is monotonically increasing with $ s $ (see Fig.~\ref{fig:nspin-10-mingap-c} for $ N = 10 $; the behavior of the gap is qualitatively similar for $ N = 8 $). In addition, for this choice of $ \tau $ and $ \mingap $ the dynamics are adiabatic and the system stays in the ferromagnetic ground state for all $ s $.

In contrast, for $ c < 1 $ the minimum gap becomes smaller than the thermal energy and thermal processes become important. In particular, we see that for $ c = 7/8 $ the success probability in the closed system setting is very close to one as the dynamics are still adiabatic, but it is evidently reduced due to the effect of independent dephasing. This effect is even more visible for $ c = 6/8 $, where dynamics are less adiabatic and the solid line (open system) is below the dotted line (closed system) along the entire evolution. Therefore, we conclude that for the specific parameter choices made here, i.\,e., in the adiabatic regime, the environment is detrimental for ARA for all the initial states we have specified.

\subsubsection{ $N = 4 $, $ \Gamma/J = 0.3 $, $ J\tau = 2500$} 
\begin{figure}[t]
	\centering
	\includegraphics[width = 0.9\columnwidth]{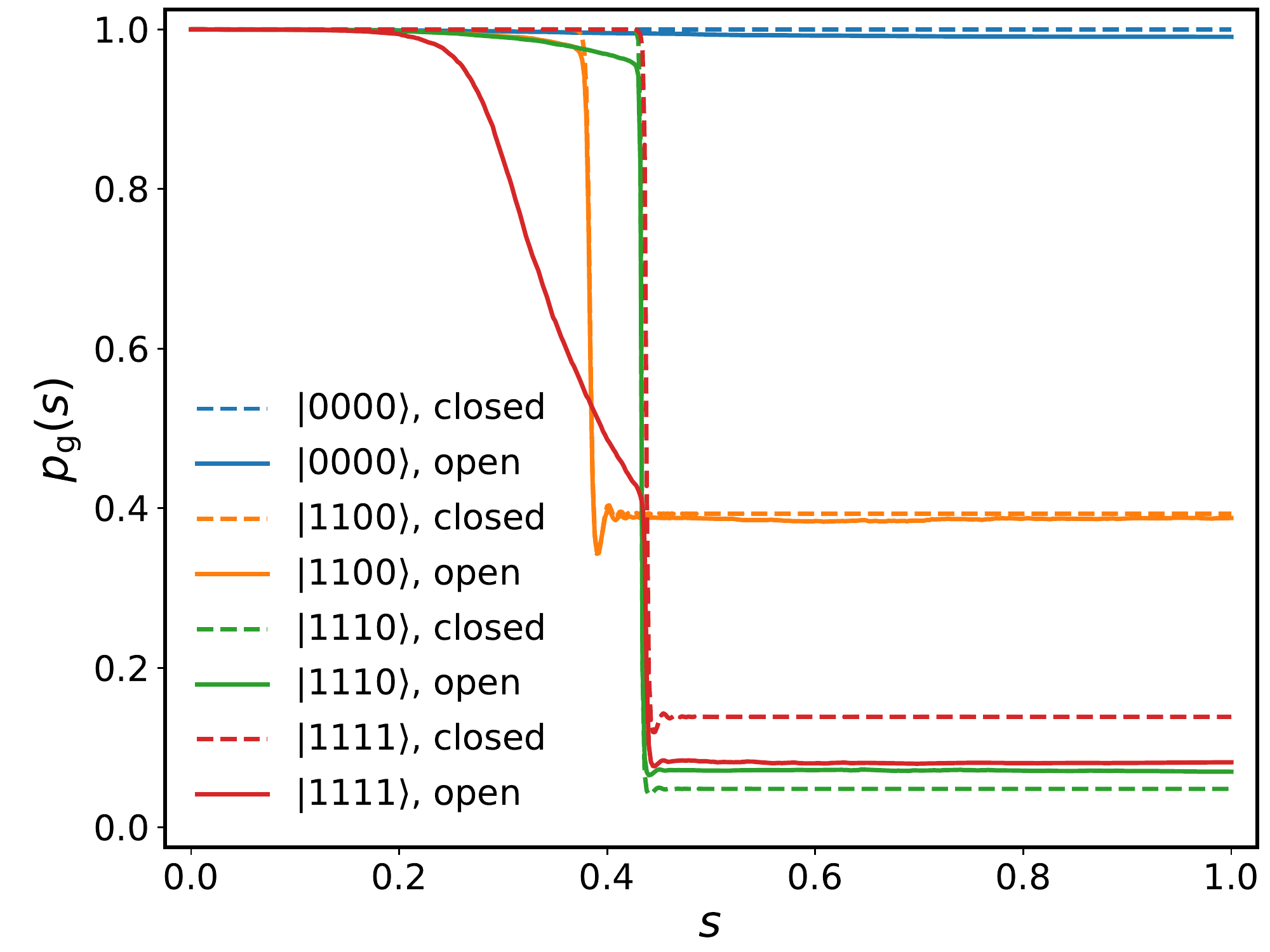}
	\caption{Open system and closed system ARA simulation results. Parameters are $ N = 4 $, $ J\tau = 2500 $, $ \Gamma/J = 0.3 $.}
	\label{fig:4ame_724_2}
\end{figure}

For this choice of parameters and for every $ c < 1 $, the dynamics are not adiabatic and the gap $ \Delta_{10} $ is very sharp around the avoided crossing. Thus, the diabatic transitions occur in a very narrow region around $ s = s_\text{min} $. In Fig.~\ref{fig:4ame_724_2}, we plot the ground state probability as a function of $ s $ for several initial states with $ c = \text{\numlist{1;2/4;1/4;0}} $. We stress that cases with $ c < 1/2 $ are very unfavorable as even a random guess of the correct solution would lead to an initial state having $ c = 1/2 $ on average in the large $N$ limit.
From Fig.~\ref{fig:4ame_724_2}, we observe sudden diabatic transitions, since $\Gamma/J = 0.3$ results in very small gaps for most of the $H_{\text{init}} $. Meanwhile, a relatively long (compared to thermal relaxation rates) annealing time of $ J\tau = 2500 $ allows for open system relaxation mechanisms to increase the instantaneous and final success probability for {one} of the initial states. Thus, in the nonadiabatic regime, $\mathsf{ARA_\text{Open}}$ can yield a higher success probability than $\mathsf{ARA_\text{Closed}}$.

\section{Time to solution}\label{sec:tts}
\subsection{Collective dephasing}

\begin{figure*}[t]
	\centering
	\subfigure[\ ]{\includegraphics[width = \columnwidth] {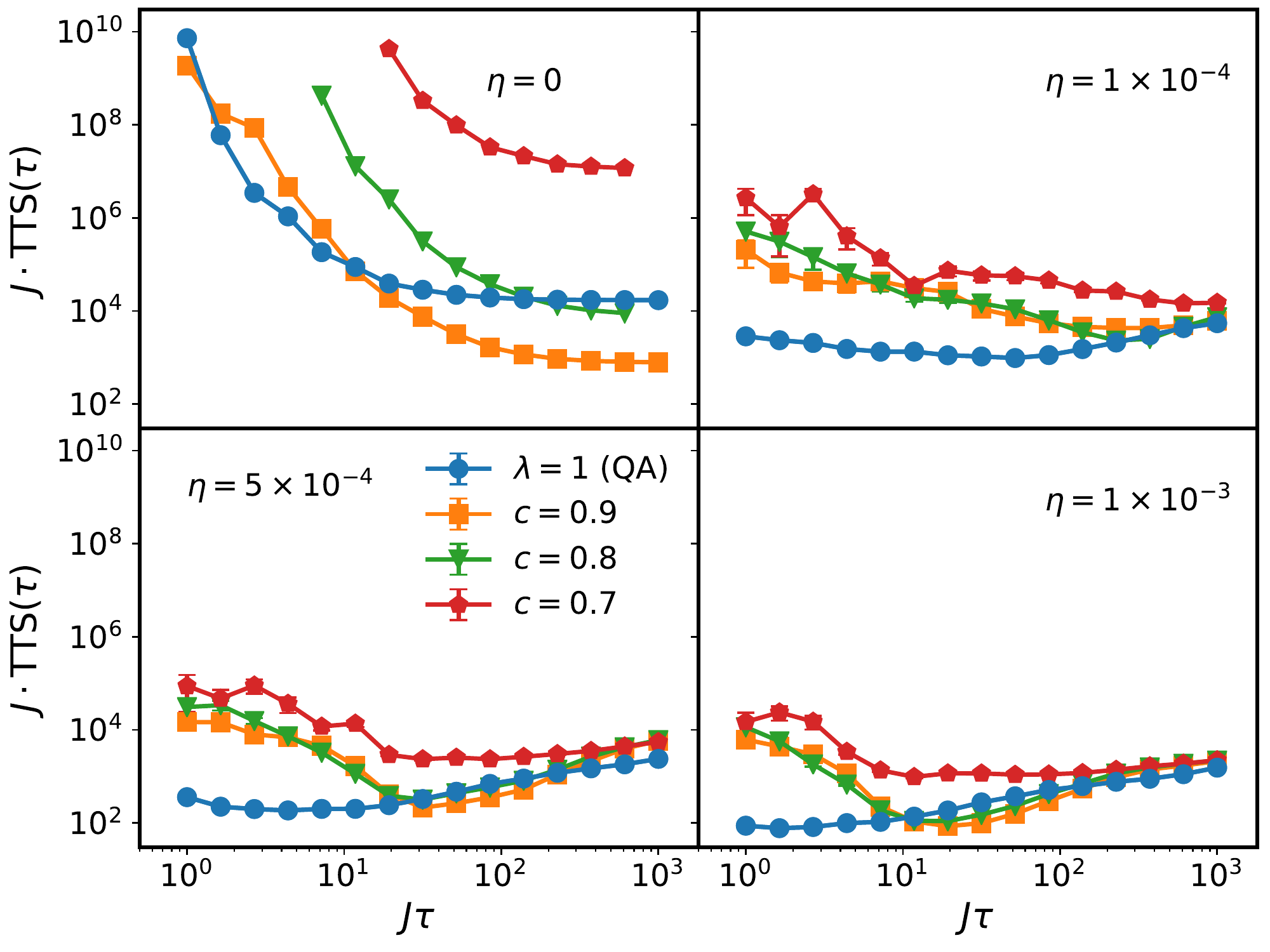}\label{fig:tts-gamma-1}}
	\subfigure[\ ]{\includegraphics[width = \columnwidth] 	{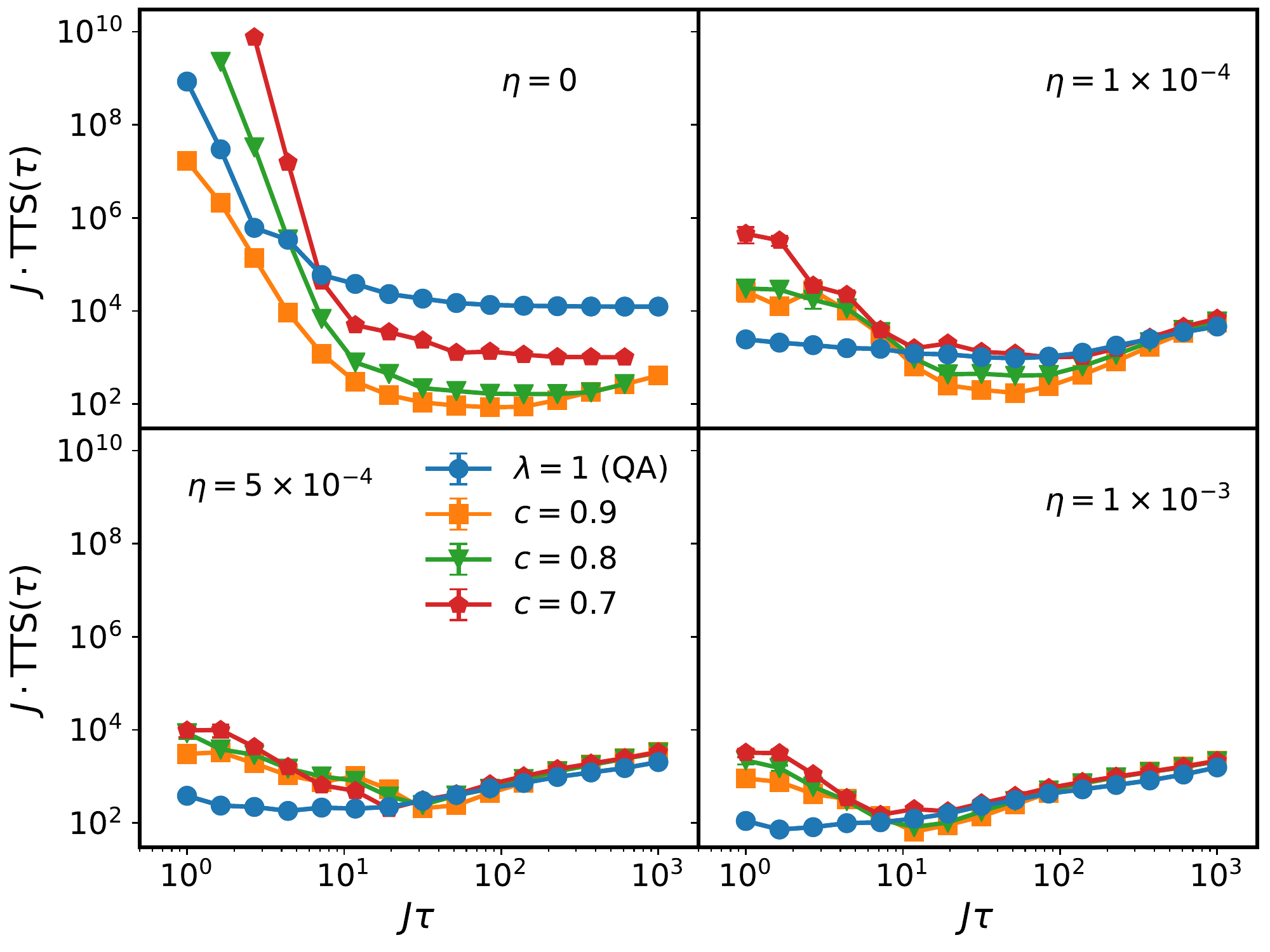}\label{fig:tts-gamma-2}}
	\caption{TTS for the collective dephasing model as a function of the annealing time $ \tau $, for several choices of the initial spin-up fraction $ c $ and for $ N = 45 $. (a) $ \Gamma/J = 1 $, (b) $\Gamma/J = 2$. Top-left: $ \eta = 0 $ (closed system case); top-right: $ \eta = \num{1e-4} $; bottom-left: $ \eta = \num{5e-4} $; bottom-right: $\eta = \num{1e-3}$.}
	\label{fig:tts-gamma}
\end{figure*}

%\DL{In the next subsection we have ``The bath is in equilibrium at temperature $ T =  \SI{12}{\milli\kelvin} $. The cutoff frequency is $ \omega_\text{c} = 8\pi $ in units of $ J $.'' What are the parameters here?}
In this section we compare ARA and QA (in both the closed and open system settings) in terms of the time to solution. To this end, we first study a system of $ N = 45 $ qubits with collective dephasing and work in the symmetry subspaces with maximal total spin. For annealing times $ J\tau \in [1, 1000] $, we compute the TTS for several values of the system-bath coupling strength $ \eta $ and for several values of $c$, the initial fraction of spin-up qubits. The bath is in equilibrium at temperature $ T =  \SI{12}{\milli\kelvin} $. The cutoff frequency is $ \omega_\text{c} = 8\pi \, \si{\giga\hertz} $. In Fig.~\ref{fig:tts-gamma-1}, we report our results for a transverse field strength of $ \Gamma/J = 1 $. For $ \eta \ne 0 $, the error bars correspond to standard wave function Monte Carlo errors (see Refs.~\cite{yip:mcwf,passarelli:pausing}) and are smaller than the point size in many cases. We can distinguish three different behaviors at short, intermediate, and long annealing times.  

For $ \eta = 0 $ (\ie, $ \mathsf{ARA}_\text{Closed} $ and $ \mathsf{QA}_\text{Closed} $, top left panel), the time to solution is a monotonically decreasing function of $ \tau $. For short $ \tau $, the success probability of both ARA and standard QA is small, hence the TTS is very large due to vanishingly small denominators in Eq.~\eqref{eq:tts}. For intermediate annealing times, the fidelity grows; the TTS first decreases in the LZ regime, and eventually saturates to a plateau as discussed in Sec.~\ref{sec:dynamics}. Longer annealing times would yield a $ \tau/\log(\tau) $ dependence of the TTS, but this region is beyond the range of annealing times analyzed in this case. These results are also thoroughly discussed in Ref.~\cite{yamashiro:ara}.	

Conversely, for $ \eta \ne 0 $ (\ie, $ \mathsf{ARA}_\text{Open} $ and $ \mathsf{QA}_\text{Open} $) some of the curves show a non-monotonic behavior, more evident for larger values of $ \eta $. At short times (\ie, in the non-adiabatic regime), the environment is beneficial in all analyzed cases and the TTS is reduced compared to $ \eta = 0 $, in agreement with many previous findings reporting the enhancement of the success probability of (several kinds of) quantum annealing of the $p$-spin model with collective dephasing~\cite{passarelli:pspin,passarelli:pausing,passarelli:reverse-ira}. At intermediate times, the effect of the environment is still generally beneficial for the TTS. Open system dynamics yield a TTS that is comparable with the closed system case in all the instances we analyzed. For $ c = \text{\numlist{0.8;0.9}} $ the curves for larger values of $ \eta $ exhibit a minimum, corresponding to an optimal working point of open system quantum annealing~\cite{lidar:open-systems,arceci:owp}, as a result of a compromise between adiabaticity and decoherence~\cite{PhysRevA.94.042131}. The behavior at long times depends on the value of $ c $. If $ c \lesssim 0.8 $, the TTS in the open system case is reduced compared to the closed system case. This is another indication of a non-adiabatic regime due to the fact that these values of $ c $ do not allow avoiding the critical point of the $p$-spin model~\cite{yamashiro:ara}. Instead, if $ c = 0.9 $, the open system TTS is larger than the closed system one for the same value of $c$ and for sufficiently large values of $J\tau$ ($\gtrsim 10^2$). When $ c = 0.9 $ the initial state is already close to the target ground state, hence the TTS is already very short in the closed system case and is harmed by decoherence. Concerning standard QA, the open system TTS is always shorter than the closed system TTS in the time window we have analyzed.

\begin{figure*}[t]
	\centering
	\subfigure[\ ]{\includegraphics[width = \columnwidth] {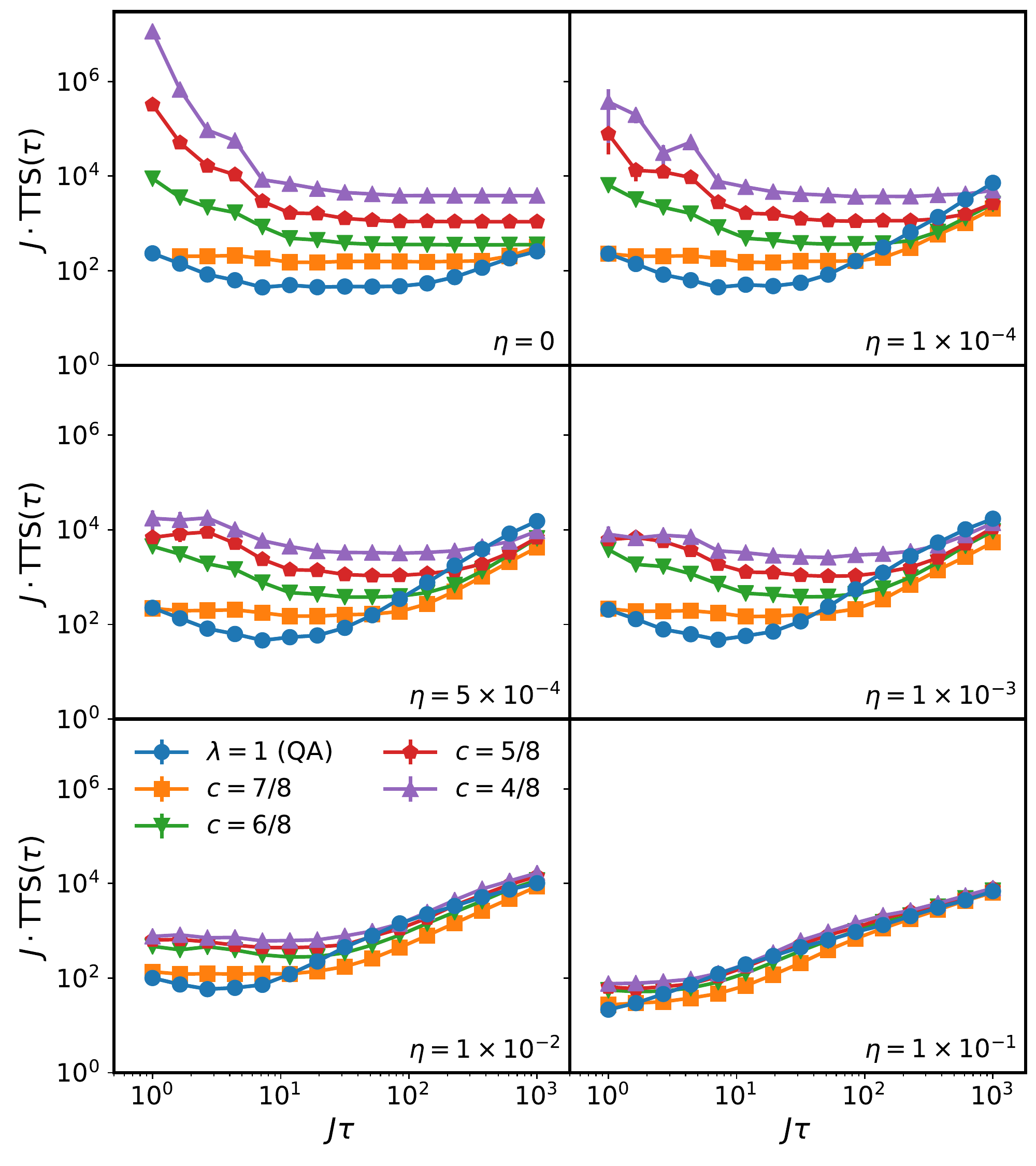}\label{fig:tts-gamma-1-n-8}}
    \subfigure[\ ]{\includegraphics[width = \columnwidth] {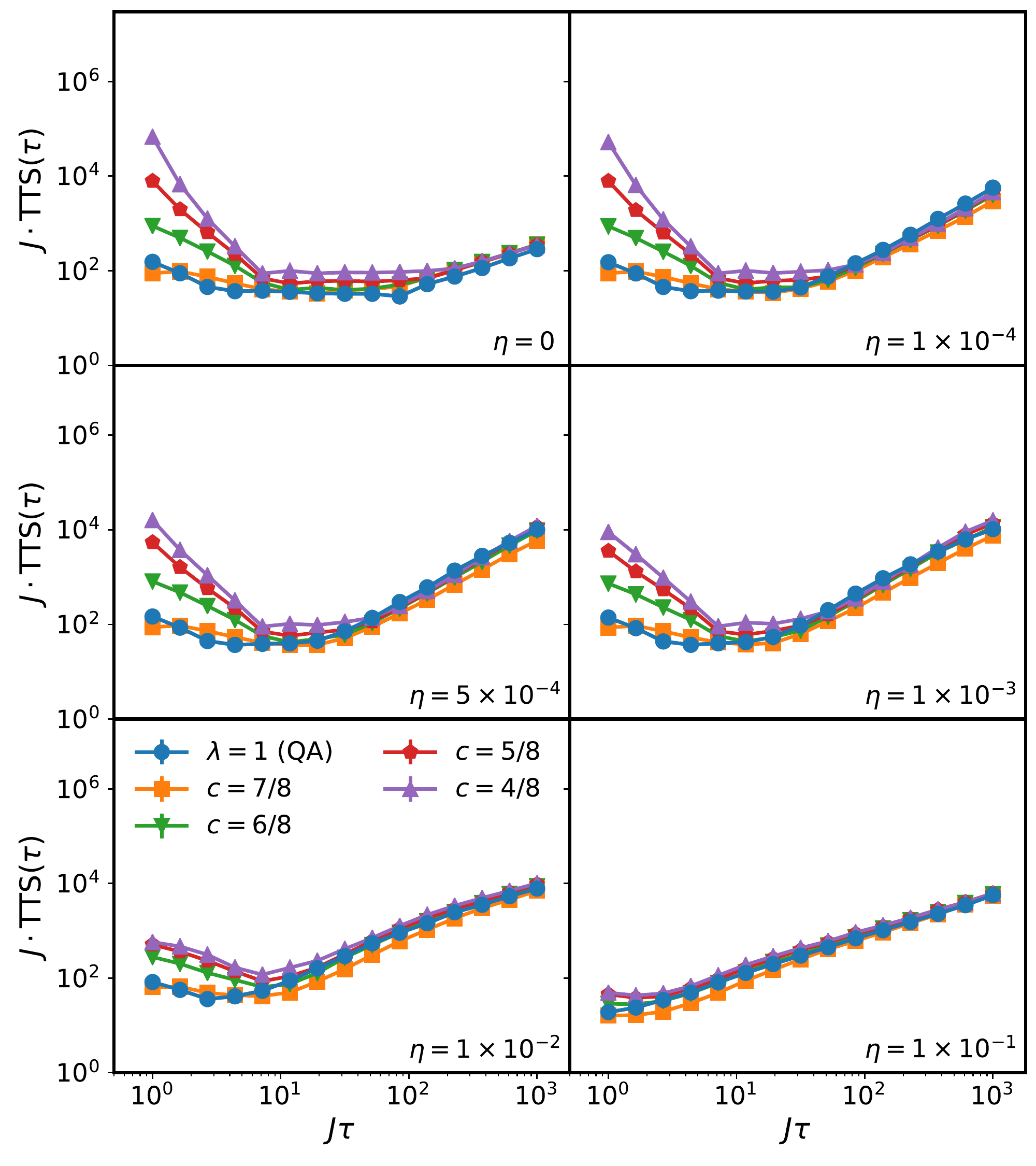}\label{fig:tts-gamma-2-n-8}}
	\caption{TTS for the independent dephasing model as a function of the annealing time $ \tau $, for several choices of the initial spin-up fraction $ c $ and for $ N = 8 $. (a) $ \Gamma/J = 1 $, (b) $\Gamma/J = 2$. Top-left to bottom-right: increasing values of $\eta$ as indicated in the legends, starting from $ \eta = 0 $ (closed system case), to $\eta = \num{1e-1}$.}
	\label{fig:tts-gamma-n-8}
\end{figure*}

As a general trend, we see that for $ \eta > 0 $ the TTS for standard QA is shorter than that of ARA both at short and long annealing times. In contrast, at intermediate times the TTS at the optimal working point in $ \mathsf{ARA}_\text{Open} $ for $ c \gtrsim 0.8 $ is shorter than the TTS of standard $ \mathsf{QA}_\text{Open} $ when $ \eta > \num{1e-4} $. In addition, we notice that the striking dependence of ARA's performance on the initial state tends to be lost in the open system setting. In fact, we notice that all curves relative to different values of $ c $ tend to converge as the system-environment coupling strength $ \eta $ increases. This result seems to suggest that ARA might not be as effective in realistic settings where the dynamics are not unitary, since the exponential speedup over standard QA provided by the avoidance of the 1QPT is mitigated by the presence of the environment, which tends to flatten the TTS curves  irrespective of the initial state. 

%%DL: Moved a discussion of this point to the conclusions
%Moreover, we stress that these results only apply in the weak-coupling regime, where the Markovian adiabatic master equation picture is valid. As pointed out in Ref.~\cite{bando2021breakdown}, mechanisms beyond the weak-coupling limit and involving bath spin-polarization, as the ones captured by the polaron-transformed Redfield equation~\cite{DazhiXu:110308} that are known to correctly model physical quantum annealers~\cite{dwave:nature2013,smirnov:hybrid-noise}, may drastically change this scenario. Assuming the bath can keep memory of the spin initial state for long times due to the low-frequency part of its correlation spectrum, the hypothesis that the results at the end of $ \mathsf{ARA}_\text{Open} $ may still retain a dependence on the initial state is not far fetched. We reserve this analysis to future works.

In order to further investigate the matter, we repeated our simulation for a transverse field strength of $ \Gamma/J = 2$. Our results are shown in Fig.~\ref{fig:tts-gamma-2}, where we report the TTS as a function of the annealing time $ \tau $. For $ \mathsf{ARA}_\text{Closed} $, we know from Ref.~\cite{yamashiro:ara} that increasing the transverse field strength from $ \Gamma/J = 1 $ to $ \Gamma/J = 2 $ causes the appearance of a minimum TTS at intermediate annealing times when the crossover between the LZ and the adiabatic regimes occurs. This feature survives also in the presence of decoherence, with the minimum being slightly lower as the system-bath coupling strength increases. For short and intermediate annealing times (up to $ J\tau \sim \text{\numrange{50}{100}} $), we see that decoherence is beneficial for ARA. On the other hand, for longer annealing times decoherence increases the TTS compared to the isolated case. For $ J\tau = 1000 $, the TTS decreases as $ \eta $ increases. Conversely, by decreasing the coupling strength, the TTS decreases towards the closed system limit, hence its behavior is non-monotonic. In the case of standard QA, decoherence in the presence of collective dephasing always reduces the TTS compared to the isolated case. All curves are very similar to the case $ \Gamma/J = 1 $ of Fig.~\ref{fig:tts-gamma-1}. In the closed system case, this was already reported in Ref.~\cite{yamashiro:ara}. In the open system case, these similarities are not surprising as changing the transverse field mostly affects the 
%beginning part of the dynamics, i.e., before 
position of the avoided crossing, 
%where the effect of decoherence is negligible. 
but not the magnitude of the gap.

The grouping of curves by $ \eta $ clearly shows that $ \mathsf{ARA}_\text{Closed} $ outperforms $ \mathsf{QA}_\text{Closed} $ at intermediate and long annealing times, but decoherence changes this feature. In fact, we see that $ \mathsf{QA}_\text{Open} $ always seems to outperform $ \mathsf{ARA}_\text{Open} $ at short and long annealing times. $ \mathsf{ARA}_\text{Open} $  only outperforms $ \mathsf{QA}_\text{Open} $ at intermediate annealing times, for $ c \gtrsim 0.8 $. This improvement can even be of one order of magnitude in specific cases, such as for $ c = 0.9 $, $ \eta = \num{1e-4} $. Nevertheless, also in this case we observe little dependence of $ \mathsf{ARA}_\text{Open} $'s performance on the initial state. The advantage that ARA has in a closed system setting compared to QA is thus almost entirely lost in the presence of collective dephasing. The caveat is that this conclusion is drawn on the basis of a single system size ($N=45$), and moreover it assumes the validity of the weak-coupling limit, which has been shown to break down for the $p=2$ $p$-spin model in recent experiments~\cite{bando2021breakdown}.

\subsection{Independent dephasing}

Finally, we turn our attention to $ \mathsf{ARA}_\text{Open} $ with independent dephasing. Since independent dephasing breaks the rotational invariance of the $p$-spin model and we have to work in the whole Hilbert space, here we focus on a smaller system of $ N = 8 $ qubits for numerical convenience. We assume that the qubits are coupled to identical baths with the same system-environment coupling strength $ \eta $. The bath is in equilibrium at temperature $ T =  \SI{12}{\milli\kelvin} $. The cutoff frequency is $ \omega_\text{c} = 8\pi \, \si{\giga\hertz} $. We consider as initial states for ARA the states (in the $ \sigma^z $ eigenbasis) with the first $ N_\uparrow $ spins up, followed by $ N - N_\uparrow $ spins down. We 
%consider $ N_\uparrow $ in the set $ \{7, 6, 5, 4\} $ so that 
choose $ c \in \{7/8, 6/8, 5/8, 4/8 \} $. As in the previous section, we compute the TTS for annealing times in the interval $ J\tau \in [1, 1000] $ and $ \eta \in \{10^{-4},5\times 10^{-4},10^{-3},10^{-2},10^{-1}\}$.

In Fig.~\ref{fig:tts-gamma-1-n-8}, we plot the TTS for $ \Gamma / J = 1 $. In the unitary case $ \eta = 0 $, we immediately notice that the TTS for $ \mathsf{ARA}_\text{Closed} $ is always worse than that of $ \mathsf{QA}_\text{Closed} $ in the Landau-Zener plateau region. Hints about the adiabatic regime can be seen from the rightmost part of the panel, where TTS curves saturate to the same $ \tau / \log\tau $ behavior. The onset of the adiabatic regime depends on the adiabatic time scale and thus on the initial spin-up fraction $c$, which is why this regime can only be observed for $ c = 7/8 $ in ARA and for standard QA in the window of annealing times we have considered. In particular, we remark that the $ \tau / \log\tau $ behavior is observed when $ \tau \gtrsim 15 \, \tau_\text{ad} $ in all cases. However, the presence of an environment seems to shift this onset to shorter annealing times as seen from the remaining panels. In addition, the scaling law of the TTS is no longer $ \tau / \log\tau $ in this decohered long-time regime. %A simple model that is able to capture this scaling is $ p_\text{e} (\tau) = 1 - [p_\text{T} + (p_\text{g} - p_\text{T}) \exp(-\tau / T_1)] $ where $ p_\text{T} $ is the Boltzmann thermal probability (at $ t = \tau $) at an effective temperature $ \beta^* $, $T_1$ is the relaxation time and $ p_\text{g} = 1-(\alpha \tau / \tau_\text{ad})^{-2} $. If the relaxation time is infinite, we recover the unitary adiabatic scaling of the error probability, while for very short relaxation times the system state at the end of the anneal is thermal. For $\eta = \num{1e-4}$, the curves saturate to this effective model with $ J T_1 = 500 $ and $ \beta^* = \beta / 4 $ (fitting parameters). For $ \eta = \num{5e-4} $, the relaxation time is $ J T_1 = 300 $, while for $ \eta = \num{1e-3} $, we get $ JT_1 = 100$.

Despite the fact that we are considering a different dephasing model, we observe a strong similarity between these results and those reported in Fig.~\ref{fig:tts-gamma-1}. In particular, we see that by increasing the decoherence strength the TTS curves relative to different initial states tend to collapse onto each other for intermediate and long annealing times, whereas for short annealing times they are separated and standard QA performs better than ARA. 

These general features are also present for other values of $ \Gamma $. In Fig.~\ref{fig:tts-gamma-2-n-8}, we report our results for $ \Gamma/J = 2 $. (We repeated our simulations also for $ \Gamma/J = 4 $ but the results are very similar to the ones shown here, hence are not included.) The same discussion of the case of $ \Gamma / J = 1 $ holds in this case as well, the only difference being that the TTS curves are already very close to each other in the unitary limit $ \eta = 0 $. Also here we see that QA outperforms ARA at short times, but then eventually all curves saturate to the same thermal behavior. A simple model that is able to capture this scaling is $ p_\text{e} (\tau) = 1 - [p_\text{T} + (p_\text{g} - p_\text{T}) \exp(-\tau / T_1)] $ where $ p_\text{T} $ is the Boltzmann thermal probability (at $ t = \tau $) at an effective temperature $ \beta^* $, $T_1$ is the relaxation time and $ p_\text{g} = 1-(\alpha \tau / \tau_\text{ad})^{-2} $. If the relaxation time is infinite, we recover the unitary adiabatic scaling of the error probability, while for very short relaxation times the system state at the end of the anneal is thermal. For $\eta = \num{1e-4}$, the curves saturate to this effective model with $ J T_1 = 1300 $ and $ \beta^* = \beta / 4 $ (fitting parameters). For $ \eta = \num{5e-4} $, the relaxation time is $ J T_1 = 300 $, while for $ \eta = \num{1e-3} $, we obtain $ JT_1 = 150$. The onset of this thermal tail depends on $ \Gamma $ and, for $ \Gamma/J = 1 $, is only observed for $\eta = \text{\numlist{1e-2;1e-1}}$ in the time window we have analyzed.

\begin{figure}[t]
	\centering
	\includegraphics[width = 0.9\columnwidth] {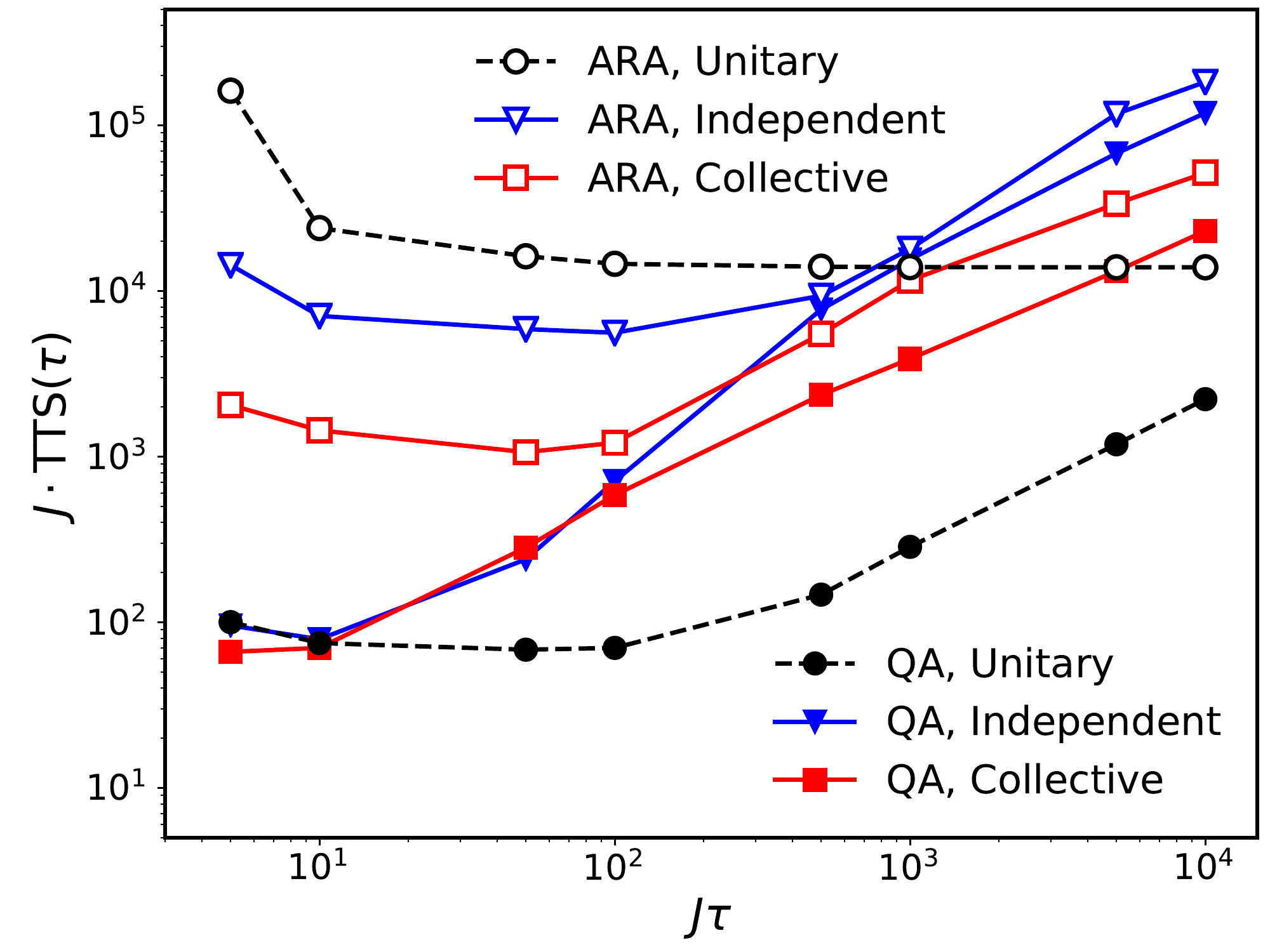}
	\caption{TTS as a function of the annealing time $ \tau $ for the random classical initial state ($ c = 0.5 $) of $ N = 10 $ qubits with $ \Gamma/J = 1 $. We compare $ \textsf{ARA}_\text{Closed} $ with $ \textsf{ARA}_\text{Open} $, $\textsf{QA}_\text{Closed}$, and $\textsf{QA}_\text{Open}$, subject to independent and collective dephasing. The coupling strength is $\eta = \num{1e-3}$.}
	\label{fig:tts-gamma-1-n-10}
\end{figure}

In order to compare the performance of $ \mathsf{ARA}_\text{Open} $ with independent and collective dephasing, we finally turn our attention to a system of $ N = 10 $ qubits and $ \Gamma/J = 1 $. The bath and coupling parameters are $T = \SI{12}{\milli\kelvin}$, $\omega_\text{c} = 8\pi \, \si{\giga\hertz} $ and $\eta = \num{1e-3}$. We consider the \textit{random} initial state with $ c = 0.5 $ and compute the time to solution as a function of $ \tau $. Our results are summarized in Fig.~\ref{fig:tts-gamma-1-n-10}. We immediately see that, in the range of annealing times we have considered, collective dephasing yields a shorter TTS compared to independent decoherence for ARA. 
In addition, we observe that, for this choice of parameters, $\mathsf{ARA}_\text{Open}$ largely outperforms $\mathsf{ARA}_\text{Closed}$ for both forms of decoherence. Finally, for both  $ \mathsf{ARA}_\text{Closed} $ and $ \mathsf{ARA}_\text{Open} $ we observe an optimal TTS. In comparing ARA to QA, it is clear that QA overall outperforms ARA, though for the independent dephasing case ARA and QA become nearly indistinguishable for sufficiently large (and suboptimal for ARA) annealing times.

%\begin{figure}[h!]
%\includegraphics[width=0.5\textwidth]{annealsuccessprob_vscol.pdf}
%\caption{Dependence of success probability as a function of $N$ due to the two different SB coupling model. $c=0.5$, $\Gamma=1$ and $\tau = 10$ns.}
%\label{fig:indvscol_n}
%\end{figure}

\section{Conclusions}\label{sec:conclusions}

%Adiabatic reverse annealing is one of the variants of QA in which partial information concerning the solution to a problem is exploited to speed up computation. 
In ARA, the system is initialized in a state that is diagonal in the computational basis, as close to the true solution as prior knowledge allows. This allows circumventing the first-order quantum phase transition of the ferromagnetic $p$-spin model with $ p\ge 3 $, thus exponentially improving the scaling of the time to solution as a function of the number of qubits, relative to standard, forward QA~\cite{yamashiro:ara}. Those findings are valid in a closed system setting, subject to purely unitary dynamics. Actual experiments of course take place under open system conditions, and this motivated us to reexamine the conclusions regarding the advantage of ARA over standard QA. 

To this end, here we applied the weak-coupling adiabatic master equation technique to study adiabatic reverse annealing in the context of the $p$-spin model (with $p = 3$), subject to collective and independent dephasing in the energy eigenbasis. By computing the time to solution and the ground state probability, we have shown that the advantage of ARA with respect to standard QA, seen in Ref.~\cite{yamashiro:ara}, disappears in the presence of weak dephasing, at least for the admittedly restricted range of parameters we were able to explore in this work. In addition, we have shown that the performance of ARA in the presence of decoherence is independent of the initial state in the adiabatic regime and weakly dependent on the initial state in the non-adiabatic regime, as opposed to the unitary case in which the choice of a suitable initial state is clearly responsible for the success or failure of the ARA protocol compared to standard QA.

%While these early results shed doubts concerning the practical usefulness of adiabatic reverse annealing, it is possible that the 
The model of decoherence provided by the AME is 
relatively benign in that it allows for successful QA due to the weak coupling assumption, which implies decoherence in the instantaneous energy eigenbasis~\cite{albash:decoherence}. However, very recent experiments that used the D-Wave quantum annealers to simulate the $p=2$ $p$-spin model~\cite{bando2021breakdown} have resulted in closer agreement with the polaron-transformed Redfield equation (PTRE)~\cite{xu_non-canonical_2016,chen2020hoqst} than the AME. The PTRE corresponds to a stronger system-bath interaction that leads to decoherence in the computational basis, but
unlike the singular coupling limit (SCL), where decoherence is also between computational basis states and which prevents any successful form of QA~\cite{albash:decoherence}, the
PTRE is governed by a non-flat (and hence non-trivial)
spectral density.
This leaves room for open system ARA to still provide an advantage over standard QA, possibly by exploiting structure in the bath spectral density. However, it may be that stronger coupling is already sufficient by itself. A hint of this possibility can be seen by observing the progression of the $c=0.5$ curves in Fig.~\ref{fig:tts-gamma-2-n-8}. The trend is that as the coupling $\eta$ is increased, the $c=0.5$ curves approach and eventually become indistinguishable from QA for the strongest coupling we have simulated ($\eta=10^{-1}$). The case of $c=0.5$ is naturally the most interesting one for ARA, since it implies an unbiased initial condition, i.e., no foreknowledge of the solution. This progression as a function of increased coupling suggests that under a more realistic model of decoherence than the AME, i.e., the PTRE, open system ARA may eventually overtake standard QA. Since this conclusion is currently only supported by our data for $\Gamma/J=2$, but not for $\Gamma/J=1$ [see Figs.~\ref{fig:tts-gamma-1-n-8} and~\ref{fig:tts-gamma-1-n-10}], additional exploration over a wide range of parameters is required before definitive conclusions can be reached. Meanwhile,
%not entirely faithful and more accurate models could provide different results instead, %or the classical spin-vector Monte Carlo~\cite{bando2021breakdown}. We postpone the analysis of these alternative models of decoherence to future works, hoping that, in the mean time, 
upcoming experiments using the D-Wave annealers, exploiting their $h$-gain feature~\cite{dwave-site} to simulate the ARA protocol, could provide additional
valuable insights.
%experimental data against which we could benchmark our numerical results.

Finally, we expect that using error suppression methods such as quantum annealing correction~\cite{PAL:13}, in particular methods that account for the need to embed a fully connected problem such as the $p$-spin model using the available connectivity of quantum hardware~\cite{Vinci:2015jt,vinci2015nested}, will significantly reduce the effective strength of the system-environment coupling, thus providing a more coherent alternative to restoring the performance of open system ARA to the level of its closed system counterpart.

\begin{acknowledgments}
    
G.\,P. and P.\,L. acknowledge financial support and computational resources from MUR, PON “Ricerca e Innovazione 2014-2020”, under Grant No. “PIR01\_00011 - (I.Bi.S.Co.)”. G.\,P. acknowledges support by MUR-PNIR, Grant. No. CIR01\_00011 - (I.Bi.S.Co.). Computation for some of the work described in this paper was supported by Dipartimento di Farmacia, Università di Napoli Federico II. This research is based upon
work (partially) supported by the Office of the Director of National Intelligence (ODNI), Intelligence Advanced Research Projects Activity (IARPA) and the Defense Advanced Research Projects Agency (DARPA), via the U.S. Army
Research Office contract W911NF-17-C-0050. The views and conclusions contained herein are those of the authors
and should not be interpreted as necessarily representing the official policies or endorsements, either expressed or
implied, of the ODNI, IARPA, DARPA, ARO, or the U.S. Government. The U.S. Government is authorized to reproduce and distribute reprints for Governmental purposes notwithstanding any copyright annotation thereon. The
authors acknowledge the Center for Advanced Research Computing (CARC) at the University of Southern California
for providing computing resources that have contributed to the research results reported within this publication. URL:
\url{https://carc.usc.edu}.

\end{acknowledgments}

%\bibliography{biblio}

\end{document}